\documentstyle[12pt,aaspp4]{article}

\newcommand\kms{\ifmmode{~\rm km\thinspace s^{-1}}\else ~km\thinspace s$^{-1}$\fi}

\lefthead{Torres, Neuh\"auser \& Guenther}
\righthead{Spectroscopic binaries in a ROSAT sample South of Taurus}

\begin{document}

\title{Spectroscopic binaries in a sample of ROSAT X-ray sources south
of the Taurus molecular clouds\altaffilmark{1}}

\altaffiltext{1}{Some of the observations reported here were obtained
with the Multiple Mirror Telescope, a joint facility of the
Smithsonian Institution and the University of Arizona.}

\author{Guillermo Torres}

\affil{Harvard-Smithsonian Center for Astrophysics, 60 Garden St.,
Cambridge, MA 02138, USA}
\authoremail{gtorres@cfa.harvard.edu}

\author{Ralph Neuh\"auser}
\affil{Max-Planck-Institut f\"ur extraterrestrische Physik, D-85740
Garching, Germany}
\authoremail{rne@xray.mpe.mpg.de}

\author{Eike W. Guenther}
\affil{Th\"uringer Landessternwarte Tautenburg,
Karl-Schwarzschild-Observatorium, Sternwarte 5, 07778 Tautenburg,
Germany}
\authoremail{guenther@uranus.tls-tautenburg.de}

\begin{abstract}

We report the results of our radial-velocity monitoring of
spectroscopic binary systems in a sample of X-ray sources from the
ROSAT All Sky Survey south of the Taurus-Auriga star-forming region.
The original sample of $\sim$120 sources by Neuh\"auser et al.\ was
selected on the basis of their X-ray properties and the visual
magnitude of the nearest optical counterpart, in such a way as to
promote the inclusion of young objects. Roughly 20\% of those sources
have previously been confirmed to be very young.  We focus here on the
subset of the original sample that shows variable radial velocities
(43 objects), a few of which have also been flagged previously as
being young.  New spectroscopic orbits are presented for 42 of those
systems. Two of the binaries, RXJ0528.9$+$1046 and RXJ0529.3$+$1210,
are indeed weak-lined T~Tauri stars likely to be associated with the
$\lambda$~Ori region.  Most of the other binaries are active objects
of the RS~CVn-type, including several W~UMa and Algol systems.  We
detect a strong excess of short-period binaries compared to the field,
and an unusually large fraction of double-lined systems. This, along
with the overall high frequency of binaries out of the original sample
of $\sim$120 sources, can be understood as a selection effect since
all these properties tend to favor the inclusion of the objects in a
flux-limited X-ray survey such as this by making them brighter in
X-rays.  A short description of the physical properties of each binary
is provided, and a comparison with evolutionary tracks is made using
the stellar density as a distance-independent measure of evolution. We
rely for this on our new determinations of the effective temperature
and projected rotational velocities of all visible components of the
binaries.  A number of the systems merit follow-up observations,
including at least 4 confirmed or probable eclipsing binaries. One of
these, RXJ0239.1$-$1028, consists of a pair of detached K dwarfs and
may provide for a potentially important test of stellar evolution
models once the absolute dimensions of the components are determined. 
	
\end{abstract}


\section{Introduction}
\label{secintro}

With the launch of the ROSAT satellite and of its predecessor, the
Einstein Observatory, a new dimension was added to the study of
star-forming regions that has enabled us to probe the higher energies
released by young stellar objects. Many X-ray sources associated with
known classical T~Tauri stars (cTTS) or weak-line T~Tauri stars (wTTS)
were detected by these space missions in molecular cloud regions that
have traditionally been thought of as being the natural habitat of
low-mass pre-main sequence (PMS) stars. Additional wTTS also in the
immediate vicinity of the molecular gas were discovered, nearly
doubling the known population of young stars of this kind. 

But large numbers of similar objects have also turned up \emph{around}
these regions, far removed from the presently known cloud material.
Systematic studies to search for young objects over large areas of the
sky based on the (spatially unbiased) ROSAT All Sky Survey (RASS) have
been carried out by
 \markcite{a95}Alcal\'a et al.\ (1995) in the Chamaeleon region, 
\markcite{w96}Wichmann et al.\ (1996) in the central region of the Taurus-Auriga association,
\markcite{a96}Alcal\'a et al.\ (1996) in Orion, 
\markcite{k97}Krautter et al.\ (1997) in Lupus, 
\markcite{n95a}Neuh\"auser et al.\ (1995a; hereafter N95a) and
\markcite{m97}Magazz\`u et al.\ (1997; hereafter M97) south of the Taurus-Auriga region, and
\markcite{n00}Neuh\"auser et al.\ (2000) in Corona Australis.
 These studies, mostly carried out at relatively low spectral
resolution, have identified several hundred candidates that also have
all the appearance of wTTS and their sheer numbers could have a
significant impact on our understanding of issues such as the
efficiency of the process of star formation.  Follow-up studies with
high spectral resolution to confirm the characteristic spectroscopic
signatures of T~Tauri stars among these candidates (late spectral type
along with H$\alpha$ emission and strong \ion{Li}{1}~$\lambda$6708
absorption) have also been performed, and a fraction of the objects
($\sim$20-90\%, depending on the region; e.g., \markcite{c97}Covino et
al.\ 1997, \markcite{w99}Wichmann et al.\ 1999, \markcite{w00}2000)
indeed seem to be very young (1-10~Myr).  Kinematic studies to
investigate the association of these objects with their parent clouds
have provided important complementary information
(\markcite{n95a}N95a; \markcite{n97}Neuh\"auser et al.\ 1997,
hereafter N97; \markcite{a00}Alcal\'a et al.\ 2000;
\markcite{w00}Wichmann et al.\ 2000). 

Radial velocities for more than 100 X-ray sources in the region south
of the Taurus-Auriga molecular clouds were reported by
\markcite{n95a}N95a and \markcite{n97}N97. In the course of that
high-resolution spectroscopic survey we discovered several dozen
spectroscopic binaries, three of which have been identified as
pre-main sequence stars in earlier studies.  Such systems are
extremely interesting given that orbits have been determined for only
about 40 PMS systems (\markcite{m94}Mathieu 1994; \markcite{m01}Melo
et al.\ 2001), and the study of their properties can provide valuable
insights into the process of star formation.  In special cases these
binaries allow one to determine the absolute masses of the components
---the most basic property of a star--- and can serve to test models
of stellar evolution for young stars that are virtually unconstrained
by the observations so far. 

In this paper we present the results for the binary population of the
sample of X-ray sources observed by \markcite{n95a}N95a and
\markcite{n97}N97, including new orbital solutions for 42 objects and
a detailed discussion of each system.  We also report determinations
of the effective temperature and projected rotational velocity for all
visible components, which aid in establishing the nature of these
objects by comparison with recent stellar evolution models. As
expected from their detection in X-rays by ROSAT, all of them are
fairly active and this is reflected in some of their overall
properties such as the orbital period distribution, which we compare
with that for normal solar-type stars. Perhaps not surprisingly, the
great majority of the systems are most likely of the RS~CVn type, and
a few belong to the Algol or W~UMa class of interacting binaries. For
at least two of our stars the spectroscopic and dynamical evidence
suggests that they are bona-fide PMS objects, one of them possibly
being as young as 1-2~Myr. These are potentially important systems
that merit further study.

\section{The sample}
\label{secsample}

As described in more detail by \markcite{n97}N97, the objects in that
study were originally selected on the basis of a combination of their
X-ray properties as observed by ROSAT and also optical flux
information (see \markcite{s95}Sterzik et al.\ 1995). Specifically,
the X-ray hardness ratios HR1 and HR2 (\markcite{n95a}N95a), the
X-ray-to-optical flux ratios, and the visual magnitudes were used to
sieve out from among the thousands of sources detected by ROSAT the
objects that have properties similar to those of the known PMS stars,
and which would therefore also be more likely to be young (see also
\markcite{m97}M97). The sample is effectively flux-limited, the limit
being set by a maximum likelihood threshold to avoid confusion with
the X-ray background (see \markcite{n95b}Neuh\"auser et al.\ 1995b).
The 111 sources observed by \markcite{n97}N97 cover an area of several
hundred square degrees south of the Taurus-Auriga star-forming region.
Roughly 20\% of them have been identified as being PMS stars on the
basis of their spectral type and \ion{Li}{1}~$\lambda$6708 absorption. 

Of this sample of 111 stars, 40 objects are binaries for which we have
determined the spectroscopic orbits.  Ten additional sources in the
same area were added after the publication of the \markcite{n97}N97
results, and were selected using somewhat relaxed criteria (see
\markcite{z98}Zickgraf et al.\ 1998) and also proper motion
information (Frink 1997, priv.\ comm.).  Three of those are also
binaries with orbits.  A few systems flagged by \markcite{n97}N97 as
having variable or possibly variable radial velocities have not been
confirmed to vary with further observation, while others remain as
possible binaries but require continued monitoring at much higher
signal-to-noise. 

\begin{table}
\dummytable\label{tabsample}
\end{table}
\placetable{tabsample}

X-ray and optical information for the 43 objects that turned out to be
spectroscopic binaries are available from the studies by
\markcite{m97}M97 and \markcite{n97}N97, and the relevant data for the
present study are collected in Table~\ref{tabsample} for easy
reference. Additional measurements of the strength of the
\ion{Li}{1}~$\lambda$6708 line with higher spectral resolution than in
\markcite{m97}M97 were obtained specifically for this project, and are
also listed. 

One of the objects, [30]~RXJ0441.9$+$0537\footnote{The number in
square brackets preceding the ROSAT designation here and throughout
the paper is an internal reference number only (from
Table~\ref{tabsample}), and is not intended for external use.}, is a
member of a rare class of cool Algols (see, e.g., \markcite{p92}Popper
1992), and was reported earlier by \markcite{t98}Torres, Neuh\"auser
\& Wichmann (1998). The remaining objects belong to a variety of
categories ranging from normal main-sequence binaries to evolved
systems of the RS~CVn type, and are described in more detail below. 
	
\section{Spectroscopic observations and reductions}
\label{secspec}

The observations of the ROSAT sources were obtained using a variety of
telescopes and instruments, mostly at the Harvard-Smithsonian Center
for Astrophysics (CfA), but occasionally also elsewhere. 

At CfA we used three, nearly identical echelle spectrographs on the
1.5-m Wyeth reflector at the Oak Ridge Observatory (Harvard,
Massachusetts), the 1.5-m Tillinghast reflector at the F.\ L.\ Whipple
Observatory (Mt.\ Hopkins, Arizona), and the Multiple Mirror Telescope
(also on Mt.\ Hopkins) prior to its conversion to a monolithic 6.5-m
mirror. A single echelle order centered at 5187~\AA\ was recorded using
intensified photon-counting Reticon detectors, with a spectral window
of 45~\AA. The resolving power of these observations is
$\lambda/\Delta\lambda = 35,\!000$. The signal-to-noise ratios range
from about 5 to 50 per resolution element. A total of 864 usable
spectra of the stars identified as binaries in the \markcite{n97}N97
sample were obtained mostly from 1994 September to 2001 April. 

Radial velocities for the single-lined objects were obtained from
these spectra by cross-correlation using the IRAF\footnote{IRAF is
distributed by the National Optical Astronomy Observatories, which is
operated by the Association of Universities for Research in Astronomy,
Inc., under contract with the National Science Foundation.} task XCSAO
(\markcite{km98}Kurtz \& Mink 1998). Templates were selected from a
large library of synthetic spectra based on model atmospheres by R.\
L.\ Kurucz\footnote{Available at {\tt http://cfaku5.harvard.edu}.},
computed for us by Jon Morse (Morse \& Kurucz, in preparation; see
also \markcite{n94}Nordstr\"om et al.\ 1994). These calculated spectra
are available for a wide range of effective temperatures ($T_{\rm
eff}$), projected rotational velocities ($v \sin i$), surface
gravities ($\log g$) and metallicities. The optimum template for each
object was determined from extensive grids of correlations in
temperature and rotational velocity (the two parameters that affect
the radial velocities the most), for an adopted surface gravity and
for solar metallicity. Constraints on the surface gravity on a
star-by-star basis are discussed below. 

For double-lined objects radial velocities were determined using
TODCOR (\markcite{zm94}Zucker \& Mazeh 1994), a two-dimensional
cross-correlation algorithm that uses two templates, one for each
component of the binary.  In a few cases the stars were found to be
triple-lined, and in those situations we used an extension of TODCOR
to three dimensions (\markcite{ztm95}Zucker, Torres \& Mazeh 1995).

The precision of the radial-velocity measurements is typically about
0.5\kms\ for single-lined spectra with sharp lines, but can be larger
for fast rotators and for some of the double-lined objects,
particularly when the secondary components are faint. 

The zero-point of the CfA velocity system was monitored every night by
obtaining twilight exposures at dusk and dawn, and applying systematic
corrections for each observing run as described in more detail by
\markcite{l92}Latham (1992).  The accuracy of the CfA velocity system,
which is within about 0.1\kms~of the reference frame defined by minor
planets in the solar system, is documented in the previous citation
and also by \markcite{slt99}Stefanik et al.\ (1999). 

Occasional observations for a few of the stars were made also with an
echelle spectrograph on the 2-m Alfred Jensch telescope in Tautenburg
(Germany). We used the red grism covering the wavelength region from
5600~\AA\ to 9600~\AA, and with a 2\arcsec\ slit the instrument yields
a resolving power of about $\lambda/\Delta\lambda = 35,\!000$. The
signal-to-noise ratios achieved range from 30 to about 100 per pixel. 

Radial velocities from these spectra were measured using the IRAF
routine FXCOR to cross-correlate each observed spectrum against a
template. For the latter we used an observation of the late-type star
HR~5777 (37~Lib, SpT K1\thinspace III-IV), with a mean radial velocity
adopted from \markcite{m93}Murdoch, Hearnshaw \& Clark (1993).
Instrumental shifts were accounted for by monitoring the position of
the telluric O$_2$ lines. Measurements of the
\ion{Li}{1}~$\lambda$6708 line were made on the best of the Tautenburg
spectra, and are reported in Table~\ref{tabsample}. 

The measured radial velocities of all our program stars in the
heliocentric frame are listed in Table~\ref{tabsb1rvs} (for the
single-lined systems) and Table~\ref{tabsb2rvs} (double-lined
systems), available in full in electronic form. 

\section{Results}
\label{secresults}

Stellar properties ($T_{\rm eff}$, $v \sin i$) for the visible
components of all the single-lined binaries were determined from the
CfA spectra by performing cross-correlations against synthetic
templates over a broad range of values in temperature and rotational
velocity, and seeking to maximize the average correlation over all
exposures of a given object. A similar procedure was followed for the
double-lined objects using TODCOR. Because of the narrow wavelength
region covered by our observations and the relatively few spectral
lines available, the surface gravity and effective temperature are
strongly correlated so that, for example, adopting lower gravities for
our templates leads to cooler temperatures. Therefore, the value of
$\log g$ was fixed for each component of each binary based on the
nature of the star from an analysis of the orbit and other physical
characteristics. The procedure was iterated until a consistent set of
parameters was obtained. 

\begin{figure}
 \plotfiddle{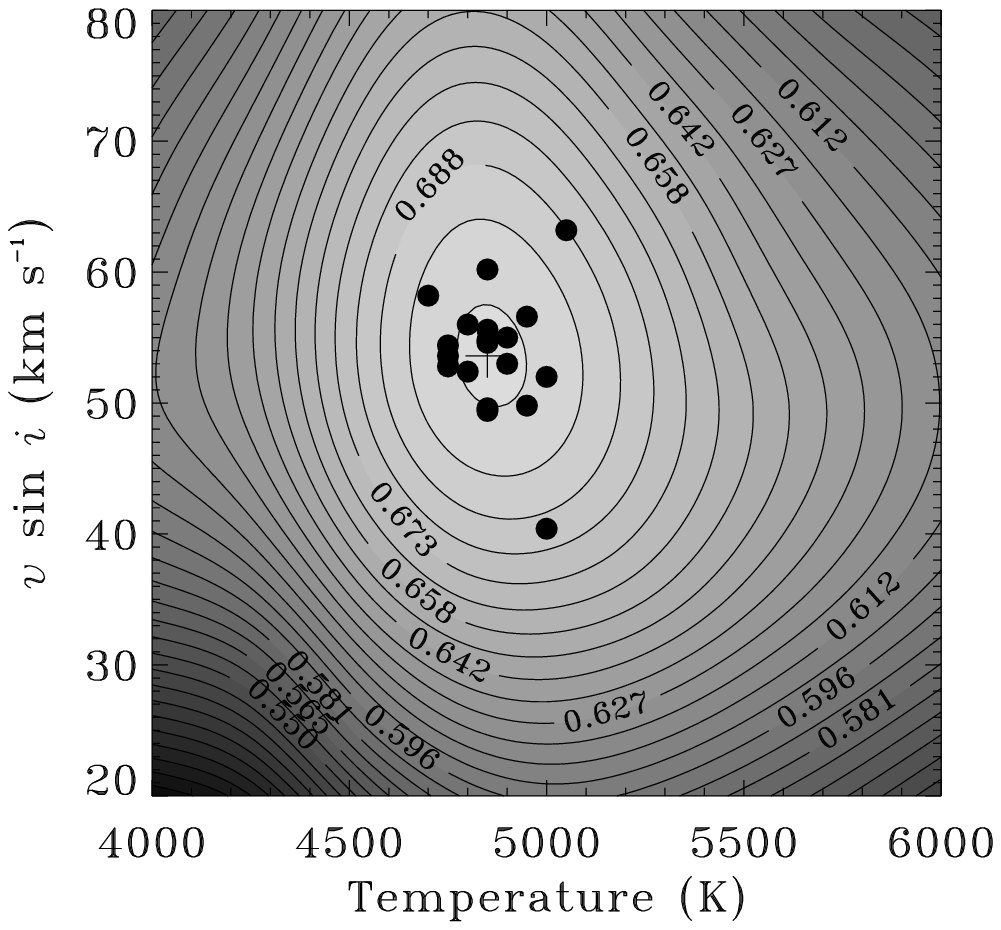}{4in}{0}{85}{85}{-260}{-170}
\figcaption[Torres.fig01.ps]{Determination of the effective temperature and
rotational velocity of the single-lined binary [15]~RXJ0340.3$+$1220.
Contours represent the correlation value averaged over all exposures
of the object, and the dots represent the parameters at the maximum
correlation for each individual exposure. The peak of the average
correlation is indicated by the plus sign ($T_{\rm eff} = 4850$~K, $v
\sin i = 54\kms$).\label{figtv1}}
 \end{figure}

\begin{figure}
 \plotfiddle{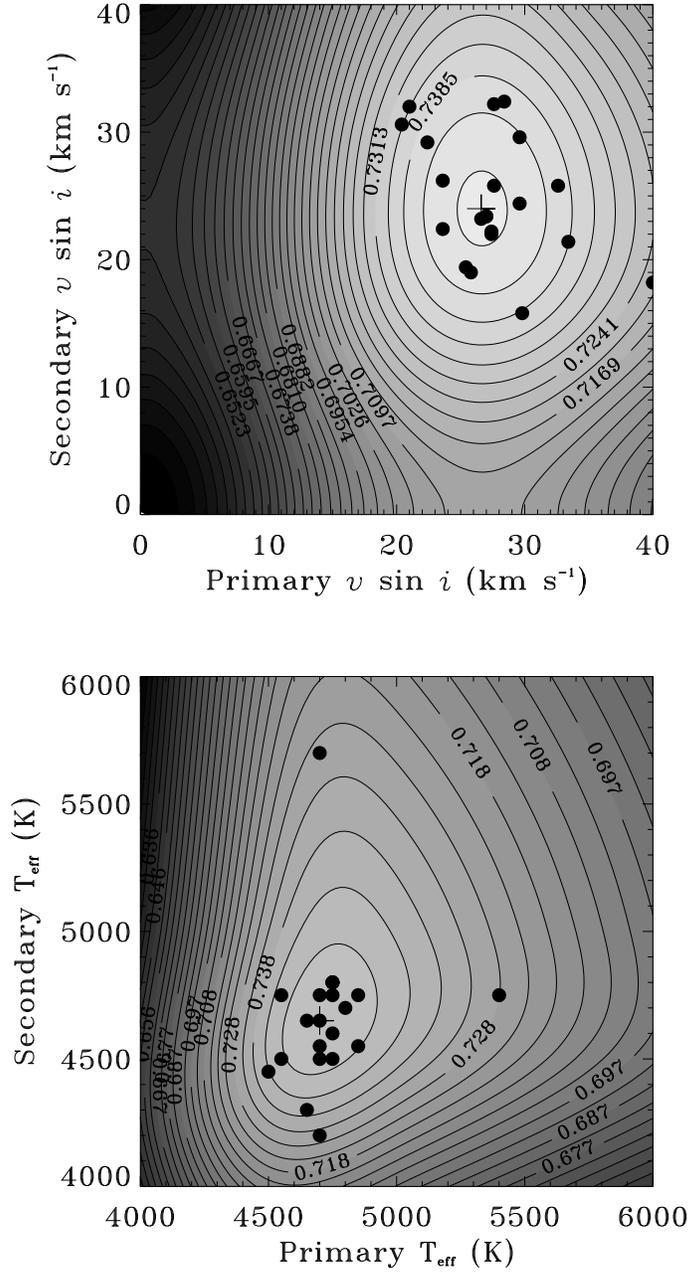}{7in}{0}{85}{85}{-260}{-70}
\figcaption[Torres.fig02.ps]{Same as Figure~\ref{figtv1}, but for the
double-lined binary [8]~RXJ0251.8$-$0203. The rotational velocities
are determined first (top panel; $v_A \sin i = 27\kms$, $v_B \sin i =
24\kms$), and then fixed in order to determine the effective
temperatures (bottom; $T^A_{\rm eff} = 4700$~K, $T^B_{\rm eff} =
4650$~K).\label{figtv2}}
 \end{figure}

An example of the $v \sin i$ and $T_{\rm eff}$ determinations for a
single-lined binary is shown in Figure~\ref{figtv1}. The contours
correspond to equal correlation value (averaged over all exposures),
and the dots represent the result for each individual exposure for the
object. For the double-lined binaries we first determined the $v \sin
i$ for the two components, which has the largest effect, and then the
temperatures for fixed values of the rotational velocity (see
Figure~\ref{figtv2}). Iterations were carried out until convergence. 
	
\placefigure{figtv1}
\placefigure{figtv2}

The effective temperatures, surface gravities, and projected
rotational velocities determined from our CfA spectra are given
separately for the single-lined binaries (SB1) in
Table~\ref{tabsb1templ} and for the double-lined binaries (SB2) in
Table~\ref{tabsb2templ}. For the latter we list also the light ratio
determined in the manner described by \markcite{zm94}Zucker \& Mazeh
(1994). The light ratios (secondary/primary) range from 0.06 to 4.1,
and although they correspond strictly to our spectral window
(5165--5211~\AA), they are quite close to the visual band. The
estimated uncertainties in the temperatures and $v \sin i$
determinations are generally $\sim$150~K and 1-3\kms, respectively,
although in some cases the errors may be considerably larger, as
indicated by a colon (:) in Tables~\ref{tabsb1templ} and
\ref{tabsb2templ}. 

\begin{table}
\dummytable\label{tabsb1templ}
\end{table}
\placetable{tabsb1templ}

\begin{table}
\dummytable\label{tabsb2templ}
\end{table}
\placetable{tabsb2templ}

\begin{figure}
 \plotfiddle{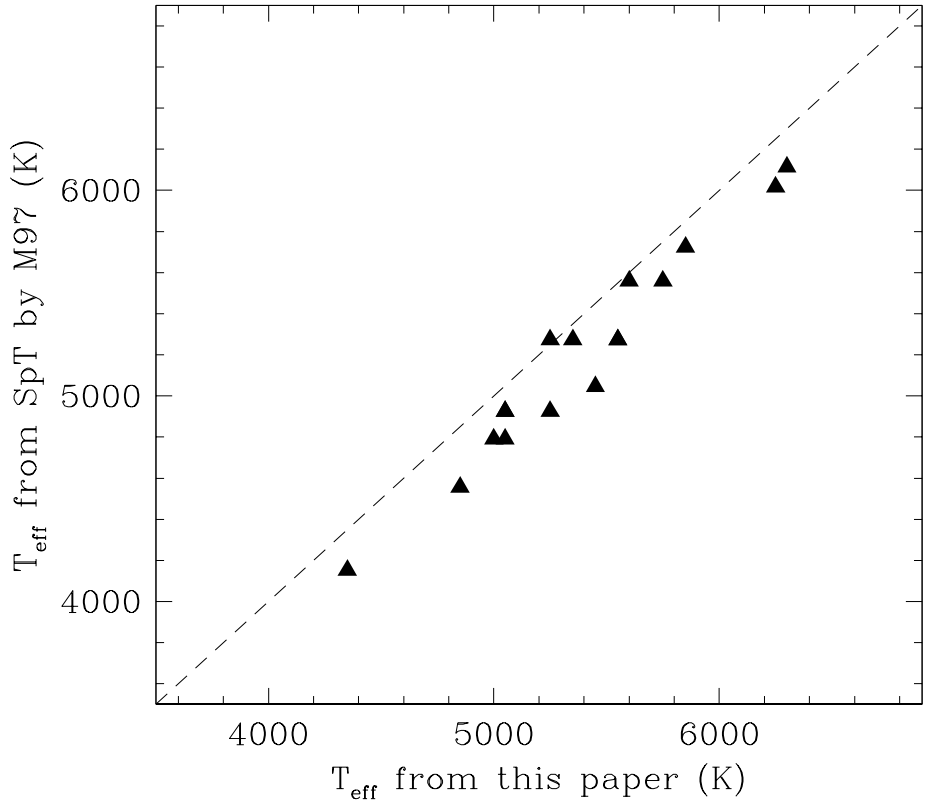}{4in}{0}{75}{75}{-230}{-120}
\figcaption[Torres.fig03.ps]{Comparison between the effective temperatures of the
single-lined binaries in our sample derived from the M97 spectral
types and the values determined here from fits to synthetic templates.
Spectral types were converted to temperatures using the tabulation by
Gray (1992). The offset of about 200~K from the diagonal line
representing a perfect correlation is clearly
seen.\label{figteffcomp}}
 \end{figure}

Spectral types for most of our stars have been published by
\markcite{m97}M97 and are collected in Table~\ref{tabsample}. Those
determinations have a quoted uncertainty of $\pm1$ or 2 subtypes,
increasing to $\pm3$ subtypes for stars earlier than about G5, mostly
due to differences in resolution between the standard library of
stellar spectra that \markcite{m97}M97 adopted for reference
(\markcite{j84}Jacoby, Hunter \& Christian 1984) and the spectroscopic
material they used (Alcal\'a 2001, priv.\ comm.).  For the
single-lined systems, where there is no confusion as to which
component the spectral type corresponds, a useful comparison with
the effective temperatures derived here is possible.  In
Figure~\ref{figteffcomp} we have converted the spectral
classifications by \markcite{m97}M97 to temperatures using the
calibration by \markcite{g92}Gray (1992). There is systematic offset
with our determinations of about $200 \pm 30$~K, our temperatures
being hotter than those indicated by the spectral types, which
suggests that one or both temperature scales could be in error.  The
difference depends, however, on the conversion table adopted. For
example, we find the offset to be $270 \pm 50$~K using the table by
\markcite{sk82}Schmidt-Kaler (1982), and $340 \pm 40$~K if we adopt
the table by \markcite{jn87}de Jager \& Nieuwenhuijzen (1987), always
in the same direction. Direct temperature determinations to test our
results based on synthetic spectra are unfortunately unavailable for
these objects.  The possibility of a systematic error in the spectral
types was investigated by checking against independent spectral
classifications by other authors. Only 6 of the stars listed by
\markcite{m97}M97 that are either single-lined binaries or single
stars (to avoid confusion in the double-lined systems) were found in
studies of ROSAT X-ray sources by \markcite{z98}Zickgraf et al.\
(1998) and \markcite{lh98}Li \& Hu (1998). Three of these objects are
among our program stars. In all six cases the spectral types by
\markcite{m97}M97 are later than the other sources, by an average of
3.3 subtypes. Though admittedly a limited comparison, the difference
is in the same sense and of the same magnitude as the discrepancy
mentioned above, suggesting that the effect may be real. We therefore
rely on our own temperature determinations in this paper for the
analysis below. 
	
\placefigure{figteffcomp}
		
Orbital solutions were obtained using standard non-linear
least-squares techniques (e.g., \markcite{pr92}Press et al.\ 1992).
For the double-lined systems the velocities of the primary and
secondary components often have very different precision due to the
different brightness or rotational velocity. For those cases the
relative weights were determined iteratively as part of the solution,
to achieve a reduced chi-square of unity. 

Inspection of the residuals of the Tautenburg observations from
preliminary orbits based on the more numerous CfA velocities revealed
that the zero-point offset between the two systems was negligible. The
two data sets were therefore merged. 

The resulting orbital elements for the single-lined binaries in our
sample are given in Table~\ref{tabsb1elem}. They are the period ($P$,
in days), the center-of-mass velocity ($\gamma$,$\kms$), the primary
velocity semi-amplitude ($K_A$,$\kms$), the eccentricity ($e$), the
longitude of periastron for the primary ($\omega$, degrees), and the
time of periastron passage for eccentric orbits or the time of maximum
primary velocity for circular orbits ($T$, heliocentric Julian date).
We list also the derived quantity $a_A \sin i$ (projected semimajor
axis of the primary, in units of $10^6$~km). The mass function $f(m)$
was also derived for each of the SB1 orbits. In Table~\ref{tabsb1elem}
we list the cube root of the mass function ($f^{1/3}$, in units of
M$_{\sun}$), which is the coefficient appearing in the expression $M_B
\sin i = f^{1/3}(M_A+M_B)^{2/3}$ that is more convenient for
evaluating the minimum mass of the secondary.  The time span and
number of observations are indicated as well, along with the rms
residual from the fit. 

\begin{table}
\dummytable\label{tabsb1elem}
\end{table}
\placetable{tabsb1elem}

For the double-lined binaries the elements are given in
Table~\ref{tabsb2elem}, and are the same as in Table~\ref{tabsb1elem}
with the addition of the velocity semi-amplitude for the secondary
($K_B$).  The derived quantities in this case are the mass ratio
($q\equiv M_B/M_A$), the relative semimajor axis ($a \sin i$, in units
of R$_{\sun}$), and the minimum masses for both components ($M_A
\sin^3 i$ and $M_B \sin^3 i$). Following the usual spectroscopic
convention, the more massive star in the system is referred to as the
primary (star~A). 

\begin{table}
\dummytable\label{tabsb2elem}
\end{table}
\placetable{tabsb2elem}

\begin{figure}
 \plotfiddle{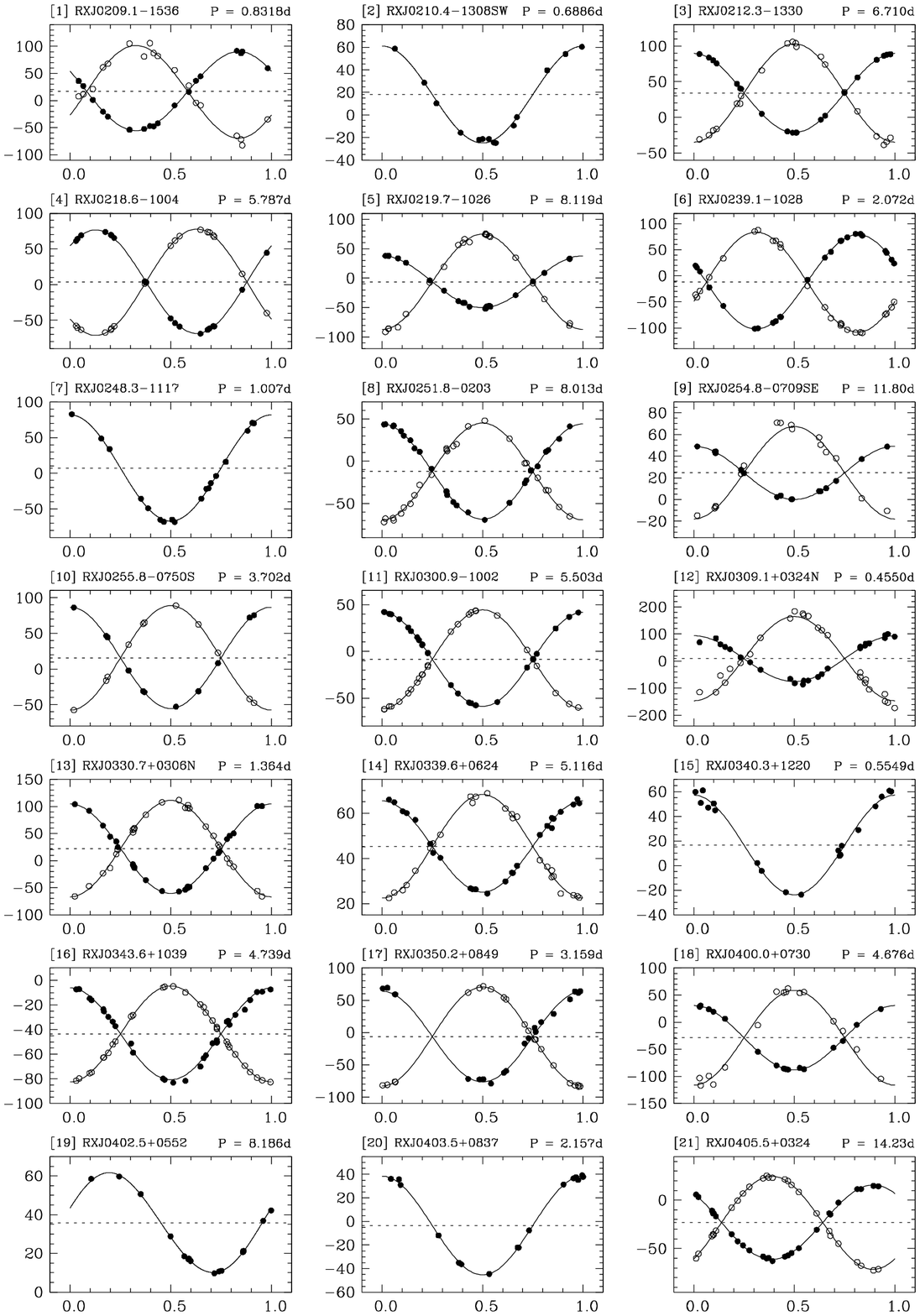}{7.34in}{0}{80}{80}{-260}{-40}
\figcaption[Torres.fig04a.ps,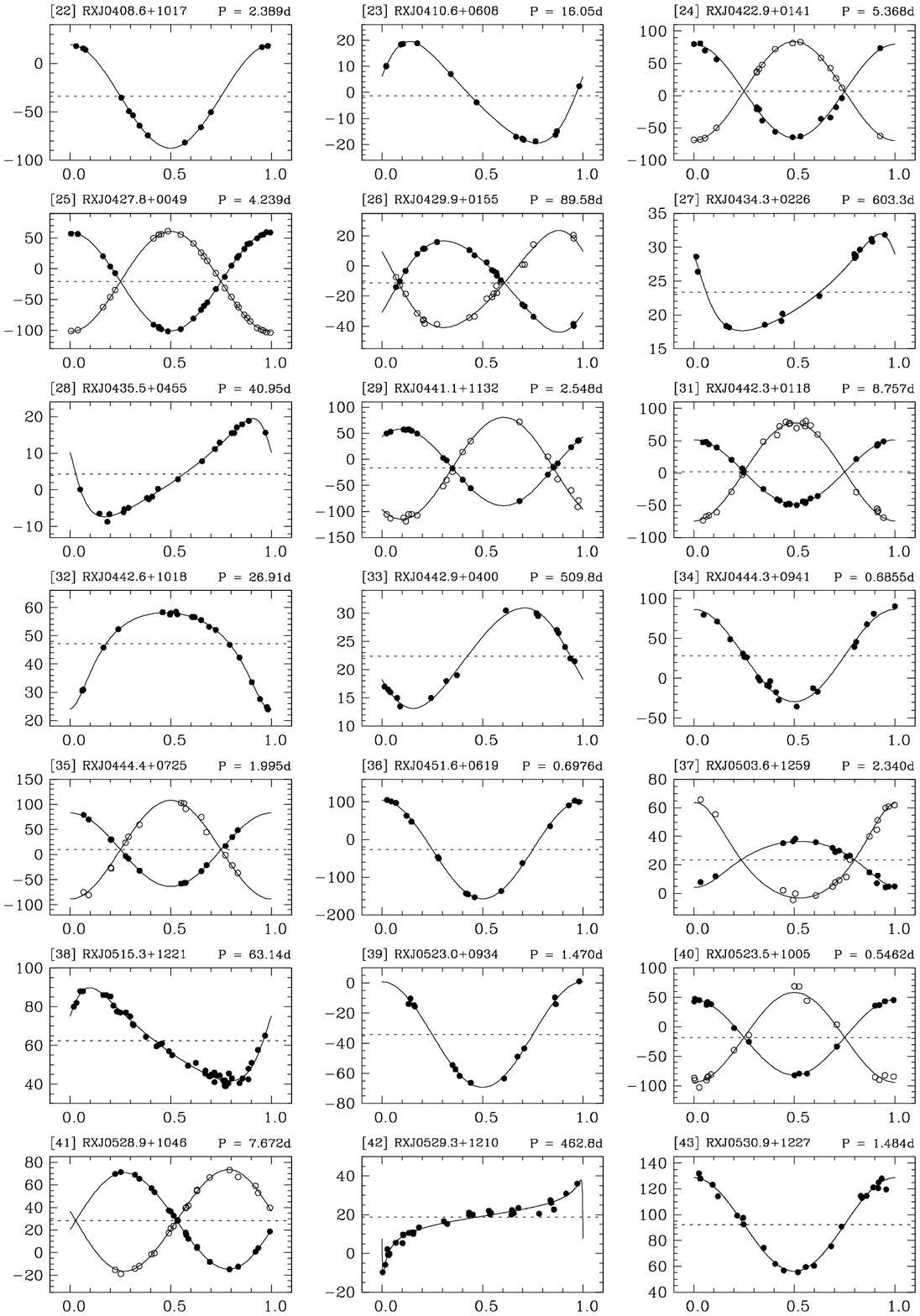]{Orbital solutions for
the single- and double-lined binaries, along with the observations.
Each plot shows the radial velocity as a function of orbital phase.
Filled symbols represent the primary velocities, and the dotted lines
indicate the center-of-mass velocity of each system. (Continued on
next page). \label{figorbits}}
 \end{figure}

The orbital solutions for all systems (except for [30] RXJ0441.9+0537;
see above) are represented graphically in Figure~\ref{figorbits}
together with the observations.  The individual residuals from the
fits are given in Table~\ref{tabsb1rvs} and Table~\ref{tabsb2rvs}. 

\placefigure{figorbits}
	
\section{Comments on individual objects}
\label{seccomments}

The orbital solutions reported in \S\ref{secresults} provide important
dynamical information on our targets that, when combined with the
measured effective temperatures and projected rotational velocities
along with other clues, offers glimpses into the nature of each of
these X-ray sources.  This, in turn, allows us to understand some of
the properties of this population of binary stars. 

In addition to comments on the particulars of some of our orbital
solutions, in this section we offer our interpretation of each system
using relatively simple hypotheses. For objects assumed to be on the
main sequence we use estimates of the mass ($M$) and radius ($R$)
based on our effective temperature determinations and standard
tabulations of average stellar properties such as that by
\markcite{g92}Gray (1992). For short-period systems, the orbits are
typically circular due to the action of tidal forces, particularly in
stars with convective envelopes. If that is the case then
synchronization between the axial rotation of the components and the
orbital motion is also a reasonable assumption, as is the alignment
between the axes of each star and the axis of the orbit. The reason
for this is that the timescales for these two processes are usually
several orders of magnitude shorter than the timescale for
circularization of the orbit, according to theory (see, e.g.,
\markcite{h81}Hut 1981). 

With the hypothesis that tidal forces have produced the
synchronization, alignment, and circularization of the orbit,
relatively robust estimates of the surface gravity for the
double-lined systems in our sample can be made based on the arguments
that follow. The minimum masses in double-lined spectroscopic binaries
are given by
 \begin{equation}
M_{A,B} \sin^3 i = 1.0361\times 10^{-1} P (K_A+K_B)^2 K_{B,A}
\end{equation}
 (in units of the solar mass when the elements are expressed in their
customary units), where the notation $M_{A,B}$ refers here to the mass
of component A or component B.  The condition of synchronous rotation
implies
 \begin{equation}
v \sin i = {50.615\over P} R \sin i~,
\end{equation}
 where $v \sin i$ is the measured projected rotational velocity in
units of$\kms$ and $P$ is the orbital period. Under the assumption of
alignment the inclination of the rotation axis ($i$) is the same as
that of the orbit.  The general expression for the surface gravity,
$\log g = 4.4377 + \log M - 2\log R$, then leads to
 \begin{equation} \log g_{A,B} = 0.8616 - \log P + 2\log(K_A+K_B) +
\log K_{B,A} - 2 \log V^{\rm rot}_{A,B} - \log \sin i~,
\end{equation} 
 in which we have abbreviated $v \sin i$ as $V^{\rm rot}$. For moderate
or large inclination angles the last term represents only a minor
correction to the surface gravity, and can be ignored (e.g., $|\log
\sin i| < 0.2$~dex for $i > 40\arcdeg$). When warranted, these
estimates of $\log g$ have been used to improve our determination of
the temperatures of the components, since the two parameters are
correlated in the cross-correlation technique we use, as mentioned
earlier.

Eq.(2) and the expression for the projected semi-major axis of the
orbit of a double-lined binary (in units of the solar radius),
 \begin{equation}
a \sin i = 1.9757\times 10^{-2} P (K_A+K_B)~,
\end{equation}
 allow one to estimate the radii of both stars in units of the
separation ($R_{A,B}/a$), independently of the inclination angle since
the $\sin i$ term cancels out. Whenever possible we have compared
these with the expected relative sizes of the Roche lobes (which
depend only on the mass ratio, $q$) to establish whether the system is
detached. Semi-detached or contact configurations are usually a sign
of mass transfer, which may contribute to the overall activity of the
systems and their strong X-ray emission. 

The relative radii also allow us to estimate the minimum inclination
angle for eclipses to occur ($i_{\rm min}$), since that angle is
simply $\cos i_{\rm min} = (R_A + R_B)/a$.  Expressed in terms of
observables, we have
 \begin{equation}
\cos i_{\rm min} = {V^{\rm rot}_A + V^{\rm rot}_B\over K_A + K_B},
\end{equation}
 where again $V^{\rm rot}_{A,B}$ and $K_{A,B}$ are in their customary
units.  Confirmation of eclipses in some of our systems would of
course provide a wealth of other information, most importantly the
knowledge of the absolute radii and masses. Very little (if any)
precise photometry is available for our targets because they are
relatively faint, but whenever available we have used the information
in the HIPPARCOS Catalogue or the Tycho-1 Catalogue
(\markcite{esa97}ESA 1997) to support our interpretation. 

As seen below, most of our systems prove to be evolved (post-main
sequence) and have overall properties that support their
classification as RS~CVn-type systems. The great majority show very
weak or no \ion{Li}{1}~$\lambda$6708 absorption in their spectra
(Table~\ref{tabsample}). In a few cases, though, the strength of the
Lithium line is strong enough that an alternate interpretation is
possible, namely, that the objects are young and are perhaps still in
the pre-main sequence phase. We discuss this evidence in more detail
in \S\ref{secpms} and \S\ref{secdensity}.  The nature of some of the
stars remains ambiguous, and follow-up observations would be useful to
improve our understanding in those cases. Summaries for each system
follow. 
	
 \noindent{\bf [1]~RXJ0209.1$-$1536.~} The secondary star in this
short-period SB2 system is relatively faint ($L_B/L_A = 0.19$), and
its parameters are somewhat uncertain. The assumption of
synchronization for the primary leads to a minimum radius of $R_A \sin
i = 0.67$~R$_{\sun}$, from eq.(2). A normal radius of 0.79~R$_{\sun}$
for a main sequence star of the measured temperature then gives $i
\sim 58\arcdeg$. However, this implies a mass for this star of only
0.30~M$_{\sun}$, which seems too small for a K0-K1 dwarf. It is more
likely that the star is larger and therefore evolved, and that the
corresponding inclination angle is smaller. From eq.(5) the condition
for eclipses is $i_{\rm min} \approx 61\arcdeg$. The system is
detached.  The formal eccentricity of the orbit is barely significant
($e = 0.0184 \pm 0.0090$, a 2$\sigma$ effect), and is probably not
real considering that the period is only 0.83~days and that both stars
are convective. 
	
 \noindent{\bf [2]~RXJ0210.4$-$1308SW.~} Single-lined binary with a
very short period and a circular orbit. If the primary has a normal
radius of about 0.75~R$_{\sun}$ for a main-sequence star of its
temperature (spectral type K2), the equatorial rotational velocity
would be very close to the value of 56\kms~that we measure for $v \sin
i$. The orbit may thus be viewed nearly edge-on.  In that case, if the
primary mass is that of a normal dwarf ($\sim0.76$~M$_{\sun}$), the
secondary would be an M5 dwarf.  Alternatively, the primary may be
somewhat evolved, and the inclination angle lower.  There is a hint of
a rising trend in the velocity residuals from the circular orbit that
needs to be confirmed. RXJ0210.4$-$1308SW is actually the fainter
component of a visual pair. The brighter star (TYC~5283-1690-1) is
$2\arcsec$ NE, and has a mean radial velocity near the center-of-mass
velocity of the object described here (\markcite{n97}N97).

 \noindent{\bf [3]~RXJ0212.3$-$1330.~} With a period of 6.7 days and a
circular orbit, the stars in this SB2 system are likely to be
synchronized with the orbital motion. The secondary is very faint
($L_B/L_A = 0.09$), and its temperature and rotational velocity are
more uncertain. The minimum mass for the primary from the
spectroscopic orbit is consistent with that expected for a normal
main-sequence star with the temperature we determine (spectral type
$\sim$K2). Similarly for the secondary. This suggests that the system
may be eclipsing, and if so, it is particularly interesting because
both stars would be under a solar mass and the system is well
detached. The entry in the Tycho catalogue (TYC~5283-876-1) indicates
a scatter in the $V_T$ passband of about 0.6~mag, although the object
is relatively faint and the significance of this somewhat questionable
in view of the relatively low precision for such objects.  Photometry
to confirm this hint is underway. The measured $v \sin i$ for the
primary star is also in agreement with the value expected from
synchronization.

 \noindent{\bf [4]~RXJ0218.6$-$1004.~} Double-lined binary with a
relatively short period (also known as TYC~5282-68-1). The stars are
expected to be synchronized, and the measured $v \sin i$ values
inserted into eq.(3) imply $\log g$ values of 3.5 and 4.0 for the
primary and secondary, respectively, which we have adopted for the
templates. The primary must therefore be evolved. This is supported by
the much lower temperature we derive for that component, which
explains the fact that the light ratio is $L_B/L_A = 2.1$ despite a
mass ratio close to unity.  The projected radii are $R_A \sin i =
3.0$~R$_{\sun}$ and $R_B \sin i = 1.8$~R$_{\sun}$, and both stars are
well within their Roche lobes.  The size of the secondary implies that
it too must be somewhat evolved. A typical mass for the temperature we
measure for that component (spectral type late F) leads to an
inclination angle of about 65$\arcdeg$. The system is thus unlikely to
be eclipsing since $i_{\rm min} \approx 73\arcdeg$.  Our orbital
solution indicates a marginal eccentricity ($e = 0.0069 \pm 0.0028$),
but this may simply be an artifact due to distortions in the
velocities related to surface activity on one or both components.

 \noindent{\bf [5]~RXJ0219.7$-$1026.~} The circular orbit in this SB2
system with a relatively short period suggests the components'
rotation may be synchronized with the orbital motion. In that case,
from the measured $v \sin i$ of the primary its minimum radius is $R_A
\sin i = 5.2$~R$_{\sun}$. Given the temperature we determine for this
star, which is 600~K cooler than the secondary, it must be a subgiant.
From eq.(3) its surface gravity is $\log g_A = 3.0$, which is the
value we used for the corresponding template.  Based on the mass ratio
this star is considerably smaller than its Roche lobe. Therefore the
system is detached. Because of the fairly extreme light ratio
($L_B/L_A = 0.06$), the rotation of the secondary is uncertain, but
definitely small. This is consistent with it being a dwarf, in
agreement with its derived temperature. We have thus adopted $\log g_B
= 4.5$ for its template. A typical mass for such a star ($M_B \sim
0.86$~M$_{\sun}$) then leads to an inclination angle of $i\sim
61\arcdeg$, and a primary mass of about 1.6~M$_{\sun}$. The system is
not eclipsing. A weak \ion{Li}{1}~$\lambda$6708 absorption line was
measured by \markcite{m97}M97 (presumably corresponding to the
brighter primary), who considered it as a possible PMS system. The Li
detection was confirmed by \markcite{n95a}N95a and \markcite{n97}N97,
but the evidence above seems to rule out the interpretation that it is
very young (see also \S\ref{secdensity}). An alternate designation for
this star is TYC~5282-2210-1.

 \noindent{\bf [6]~RXJ0239.1$-$1028.~} Another double-lined binary
with a short period and a circular orbit. The rotational velocities
expected for main-sequence stars of the temperatures we determine
(assuming radii of $\sim$0.7~R$_{\sun}$) are very close to the values
we measure, implying an almost edge-on orbit.  The expected masses are
also similar to the minimum masses we derive.  Follow-up photometry by
L.\ Marschall (Gettysburg College) has confirmed the eclipsing nature
of this system. Both stars are of spectral type mid-K, and the light
ratio is $L_B/L_A = 0.36$. Detached eclipsing binaries with relatively
low-mass components such as these are not very common, making this an
interesting system for accurate mass and radius determinations.
Photometric and spectroscopic observations are being continued towards
this goal. 

 \noindent{\bf [7]~RXJ0248.3$-$1117.~} Also known as PPM~710125. The
residuals from the circular orbit of this 1-day period SB1 system are
slightly larger than expected.  We see no sign of the secondary
component. The temperature of the primary is solar-like. If we assume
it is a main-sequence star with a radius similar to the Sun that is
rotating synchronously with the orbit, the predicted equatorial
rotational velocity is very similar to the observed $v \sin i$,
implying an orbit viewed close to edge-on. From the mass function the
secondary would then be an early M star ($M_B \sim 0.45$~M$_{\sun}$),
which would explain why it is not detected spectroscopically. If the
assumption of a main-sequence primary holds, then the system is well
detached and eclipses are possible for inclination angles larger than
$i_{\rm min} \approx 70\arcdeg$. 

 \noindent{\bf [8]~RXJ0251.8$-$0203.~} The hypothesis of
synchronization in this SB2 system with a circular orbit and a
relatively short period, along with the measured values of $v \sin i$
for both stars, lead to minimum radii of $R_A \sin i = 4.1$~R$_{\sun}$
and $R_B \sin i = 3.6$~R$_{\sun}$. These are clearly too large for the
measured temperatures, and imply that the stars are late G-type
subgiants rather than dwarfs. The surface gravities from eq.(3) are
close to 3.0 and 3.1 for the primary and secondary, respectively.  The
configuration of the system is detached, and the minimum inclination
angle for eclipses to occur would be about 64\arcdeg. The evolved
nature of the system is supported by the fact that the stars do not
quite follow the normal mass-luminosity relation for dwarfs. The
measured light ratio is $L_B/L_A = 0.71$, and from a mass ratio quite
close to unity it would be expected to be considerably
larger. \markcite{n95a}N95a reported a very weak
\ion{Li}{1}~$\lambda$6708 absorption line (equivalent width
$\sim0.06$~\AA).

 \noindent{\bf [9]~RXJ0254.8$-$0709SE.~} The physical parameters of
the stars in this rather faint double-lined binary are somewhat
uncertain. Based on grids of cross-correlations with a range of
synthetic templates it is clear that both components are very cool and
slowly rotating. Based on the dM3 classification by
\markcite{n97}Neuh\"auser et al.\ (1997) we adopt templates with a
$T_{\rm eff}$ of 3750~K and a surface gravity appropriate for dwarfs.
The mass ratio of $q = 0.577$ is, however, unusually low for a
double-lined system composed of main-sequence stars. A normal
mass-luminosity relation for dwarfs\footnote{The standard M-L relation
for dwarfs relates the mass to the \emph{bolometric} luminosity. In
this relation the exponent has a value of roughly 4. However, in the
optical bands the relation is much steeper, as described by
\markcite{c94}Carney et al.\ (1994), with exponents in the range 8-9,
depending on the band.} would imply a secondary so faint in the
visible that its lines would not be seen in the spectrum (see, e.g.,
\markcite{c94}Carney et al.\ 1994).  Yet the light ratio we measure is
$L_B/L_A = 0.19$, which suggests that the stars are perhaps not both
dwarfs.  The H$\alpha$ emission measured by \markcite{m97}M97 is the
strongest in our sample.  From the small minimum masses eclipses are
very unlikely.  A visual companion about one magnitude fainter is
located approximately $6\arcsec$ NW (\markcite{m97}M97). 

 \noindent{\bf [10]~RXJ0255.8$-$0750S.~} Double-lined binary with a
circular orbit and a 3.7-day period. The temperatures of both
components correspond to a spectral type K4 or K5.  The light ratio is
$L_2/L_1 = 0.68$. Assuming typical radii of $R \sim 0.7$~R$_{\sun}$,
synchronous rotation implies equatorial rotational velocities of
9-10\kms, marginally larger than our measured $v \sin i$ values. The
minimum masses from the orbital solution are also slightly larger than
the mass implied by the temperatures.  Adopting $M \sim
0.7$~M$_{\sun}$ we obtain $i \sim 70\arcdeg$. From the measured $v
\sin i$ values and eq.(5) the lower limit for eclipses is $i_{\rm min}
\approx 84\arcdeg$, so that eclipses are unlikely. The system is
detached, and otherwise quite similar to [6]~RXJ0239.1$-$1028. A faint
visual companion is located $10\arcsec$ north. 

 \noindent{\bf [11]~RXJ0300.9$-$1002.~} This object (also known as
HD~18775) was added to our program at a later stage. It is a
double-lined binary with a short period and a circular orbit. From the
measured values of $v \sin i$, and assuming synchronous rotation, the
surface gravities for both stars are close to $\log g = 4.0$. They are
likely to be slightly evolved.  The object was observed by the
HIPPARCOS mission (HIP~14050), and the parallax determination
(distance $=80$~pc) along with the total brightness of the pair ($V =
7.78$) and our light ratio ($L_B/L_A = 0.55$) allow the components to
be placed on the H-R diagram. Both stars lie somewhat above the main
sequence. However, the lack of \ion{Li}{1}~$\lambda$6708 absorption
(equivalent width $< 0.06$~\AA) indicates it is not young.

The system is well detached, and eclipses are possible only for
inclination angles in excess of $81\arcdeg$. Based on the minimum
masses from our spectroscopic orbit and the estimated temperatures,
the angle is almost certainly smaller than this. Eclipses are
therefore ruled out, and this is consistent with the absence of
photometric variability determined by HIPPARCOS mission. 
	
 \noindent{\bf [12]~RXJ0309.1$+$0324N.~} This SB2 has a close visual
companion approximately 2\arcsec~to the south that contaminates the
spectrum of the main target under typical observing conditions. Thus
our spectra are usually triple-lined. To handle this case we have
employed a three-dimensional cross-correlation technique (see
\markcite{ztm95}Zucker, Torres \& Mazeh 1995), which is simply an
extension of the TODCOR algorithm to three dimensions. 

The orbital period of RXJ0309.1$+$0324N (0.455 days) is the shortest
in our sample, and is the reason for the very large rotation rates of
the components.  The radial velocity precision is degraded because of
this, and the physical parameters of the secondary star are more
uncertain because in addition this component is relatively faint
($L_B/L_A = 0.24$).  The mass ratio of $q = 0.544$ is unusually low
for these to be normal stars: a standard mass-luminosity relation for
dwarfs would imply a light ratio in the visible much smaller than we
observe (\markcite{c94}Carney et al.\ 1994).  Synchronization and the
measured projected rotational velocities for the primary and secondary
lead to minimum radii of 1.06~R$_{\sun}$ and 0.85~R$_{\sun}$,
respectively. Both stars are clearly filling their Roche lobes, if not
exceeding them, making this a contact system possibly of the W~UMa
type. The mass ratio is typical for such systems, as are the
temperatures we derive.  The orbit is circular. From eq.(3) the
surface gravities are $\log g = 4.0$, ignoring the small contribution
from the unknown inclination.  The minimum inclination angle for
eclipses to occur is $\sim$27\arcdeg.  Since this angle leads to
masses that are much too large for the measured temperatures, we infer
that $i$ is considerably larger, and thus the system must be
eclipsing. The object is included in the Tycho catalog as
TYC~58-166-1. A fairly large scatter of about 0.8~mag in the $V_T$
passband is listed for this star, perhaps consistent with eclipses,
but the uncertainty in these measurements makes this evidence
inconclusive. The visual companion has a mean velocity not far from
the systemic velocity of our program star. 

 \noindent{\bf [13]~RXJ0330.7$+$0306N.~} Double-lined system
(TYC~67-206-1).  Synchronization and the measured values of $v \sin i$
imply that the primary star is approximately 1.4 times larger than the
secondary, despite the similar temperatures. The light ratio is
$L_B/L_A = 0.63$.  If the secondary is normal, its temperature
(spectral type G9) leads to a mass of $M_B = 0.86$~M$_{\sun}$, from
which the primary would be 0.92~M$_{\sun}$. The inclination angle is
then $\sim$47\arcdeg. The resulting radii (1.45~R$_{\sun}$ and
1.05~R$_{\sun}$) are both larger than normal for dwarfs, and may
indicate some degree of evolution. The system is detached, and does
not eclipse ($i_{\rm min} \approx 67\arcdeg$). A faint visual
companion is located some $30\arcsec$ south. 

 \noindent{\bf [14]~RXJ0339.6$+$0624.~} The lower temperature for the
primary compared to the secondary in this SB2 system strongly suggests
that it is an evolved star. The measured $v \sin i$ of the primary
along with the assumption of synchronous rotation, supported by the
short period and circular orbit, indicate $\log g_A = 2.27 - \log \sin
i$. Because the minimum masses are small, the inclination angle is
also likely to be small so that its contribution to the predicted
value of the surface gravity may be significant. If we make the
simplest assumption that the secondary is a normal main-sequence star,
its temperature corresponds to a mass of 1.17~M$_{\sun}$ (spectral
type F9), and the implied inclination angle is just under 15\arcdeg.
The resulting primary surface gravity is then approximately $\log g_A
= 2.9$. Independently of the assumption on the secondary, the primary
star appears to be large enough to fill its Roche lobe, indicating
mass transfer. With a radius typical for an F9 dwarf, the secondary
would be much smaller than its critical surface. A consistent picture
then emerges of a semi-detached Algol-type binary in which the mass
gainer is not as hot as in the ``classical" Algols, and the mass
transfer is probably still in the early stages such that the mass
ratio has not yet reversed to become as extreme as in those systems.
RXJ0339.6$+$0624 is unlikely to be eclipsing since the lower limit on
$i$ for this to be possible is $i_{\rm min}\approx 61\arcdeg$. The
light ratio in this system is $L_B/L_A = 0.60$. \markcite{m97}M97
measured weak \ion{Li}{1}~$\lambda$6708 absorption, but their
suggestion of a possible PMS nature is not confirmed by the dynamical
evidence presented above. See also \S\ref{secdensity}. 

 \noindent{\bf [15]~RXJ0340.3$+$1220.~} From the very short period
(0.555~days), the primary in this SB1 system cannot be a very large
star.  Assuming it is a main-sequence object, the effective
temperature leads to a mass and a radius about 0.7 times that of the
Sun, corresponding to spectral type K4.  The predicted equatorial
rotational velocity is then about 64\kms, and our $v \sin i$ of
54\kms~leads to $i\sim 57\arcdeg$.  The secondary mass derived from
the mass function is expected to be about 0.17~M$_{\sun}$,
corresponding roughly to an M4 star. With the implied mass ratio, $q
\approx 0.24$, the system is well detached.  Given the radii and
separation of the stars, eclipses are ruled out.

 \noindent{\bf [16]~RXJ0343.6$+$1039.~} With a mass ratio close to
unity and a light ratio $L_B/L_A = 4.0$, one or both stars in this SB2
must be evolved.  The fact that the primary temperature is cooler than
the secondary is also symptomatic.  From eq.(3) and the measured
values of $v \sin i$ the primary has a surface gravity about 0.6~dex
lower than the secondary, and is probably an early K subgiant. The
system is detached, and the minimum inclination angle that would allow
eclipses to occur is about 68\arcdeg. Based on the minimum masses, the
inclination angle is certainly smaller than this, so the system cannot
be eclipsing. If a normal main-sequence mass of 1.3~M$_{\sun}$ is
adopted for the secondary based on its $T_{\rm eff}$, the resulting
inclination angle is in fact $\sim$26\arcdeg.  The corresponding
radius of the secondary comes out $R_B = 1.9$~R$_{\sun}$, which is a
bit large for a main sequence star of this temperature (spectral type
mid F). It too may be slightly evolved. The surface gravities we have
adopted to derive the final temperatures are thus $\log g_A = 3.5$ and
$\log g_B = 4.0$. The presence of weak Li absorption, presumably
corresponding to the much brighter secondary, prompted
\markcite{m97}M97 to suggest a possible PMS nature for the
object. This seems unlikely (see also \S\ref{secdensity}).

The residuals of the primary star from our orbital solution are
somewhat larger than expected and show systematic patterns perhaps
indicative of surface activity (spotting). This could possibly be
related to the strong X-rays in the system. Though not listed in the
HIPPARCOS catalog, the object is included in the Tycho catalog
(TYC~660-825-1) and there are indications that it may be
photometrically variable (scatter in $V_T$ $\sim$ 0.6~mag). 

 \noindent{\bf [17]~RXJ0350.2$+$0849.~} Short-period SB2 system with a
circular orbit. The lower temperature for the more massive primary is
indicative of evolution, at least in that star, as is the fact that
the secondary is somewhat brighter ($L_B/L_A = 1.22$) for a mass ratio
under unity. This is confirmed by the minimum radius for the primary
that we derive from the measured rotational velocity and the
hypothesis of synchronization: $R_A \sin i = 3.1$~R$_{\sun}$. The
secondary also appears larger than normal for a main sequence star of
its temperature ($R_B \sin i = 1.2$~R$_{\sun}$, for SpT $\sim$ G7), so
it too must have evolved to some extent. The system is detached,
although the primary is about 86\% of the size of its critical
equipotential surface.  The minimum inclination angle for eclipses to
occur is 63\arcdeg, but given the relatively small minimum masses we
derive, eclipses seem unlikely.

 \noindent{\bf [18]~RXJ0400.0$+$0730.~} The secondary in this
double-lined system is extremely faint ($L_B/L_A = 0.10$), which
affects the precision of the radial velocities of that star and also
makes the determination of its temperature and rotational velocity
considerably more uncertain than those of the primary. Nevertheless,
it is clear that the secondary is of much earlier spectral type (mid
A?) than the more massive primary, which must therefore be evolved
(early G-type giant or subgiant).  Synchronization in the system is a
reasonable assumption for the short period and circular orbit. This
along with our measured value of $v \sin i$ for the primary leads to a
minimum radius $R_A \sin i = 3$~R$_{\sun}$, which is much larger than
normal for a star of its temperature, and is therefore consistent with
it being evolved. The system is detached. 

 \noindent{\bf [19]~RXJ0402.5$+$0552.~} If synchronization is assumed
in this 8.2-day SB1, a normal radius for a star of its temperature
leads to an equatorial velocity slightly under 6\kms, which is lower
than the measured value $v \sin i = 14\kms$.  This would suggest that
the radius is at least 2.3~R$_{\sun}$, implying that the primary star
is slightly evolved, perhaps being an early G-type subgiant.
Accordingly, we adopted a surface gravity of $\log g = 4.0$. The
formal orbital solution gives a small but perhaps significant
eccentricity of $e = 0.029 \pm 0.010$ ($2.9\sigma$). Though this may
appear to undermine the assumption that the rotation is synchronized
with the mean orbital motion, it is still quite possible that the
system is pseudo-synchronized (i.e., synchronized at periastron) if
the eccentricity is real, and the picture above would still hold.
Alternatively, the measured eccentricity could be due to velocity
distortions related to the activity in the binary.  Further
observations are needed to confirm the significance of the non-zero
eccentricity. The star is also known as TYC~79-729-1. 

 \noindent{\bf [20]~RXJ0403.5$+$0837.~} Synchronous rotation of the
primary in the short-period circular orbit of this SB1 implies a
minimum radius of $R_A \sin i = 1.45$~R$_{\sun}$, based on our
measured $v \sin i$.  Since this is larger than normal for a mid-G
type dwarf, as inferred from the temperature, we conclude that the
star is slightly evolved, and we have adopted a lower surface gravity
accordingly ($\log g = 4.0$). 

 \noindent{\bf [21]~RXJ0405.5$+$0324.~} The stars in this double-lined
system (TYC~76-713-1) are similar in brightness ($L_B/L_A = 1.11$),
but the mass ratio is smaller than unity ($q = 0.80$), and once again
the secondary (of spectral type mid F) is considerably hotter than the
primary. This suggests that the latter star is evolved (late G or
early K subgiant or giant). The rotational velocity of the secondary
is small and somewhat uncertain. A typical mass of 1.44~M$_{\sun}$ for
a star of its temperature leads to an inclination angle of roughly
42\arcdeg, and a primary mass near 1.8~M$_{\sun}$.  From the
hypothesis of synchronization and the measured rotational velocities
the surface gravity of the primary would be lower by $\Delta\log g
\approx 0.6$~dex, and its radius would be $R_A \sim 6.7$~R$_{\sun}$.
The system is detached, and a minimum inclination angle $i_{\min}
\approx 75\arcdeg$ indicates no eclipses. 

The rather long orbital period of this binary (14.2~days) and its
marginal eccentricity ($e = 0.0102 \pm 0.0048$, 2.1$\sigma$) may cast
some doubt on the validity of the assumption of synchronization. The
independent evidence (from $q$ and $L_B/L_A$) that the primary is
evolved implies, however, that this star is larger and has a deeper
convective envelope than a normal dwarf.  Tidal forces are therefore
expected to be much stronger, and possibly enough to have synchronized
the axial spin of each star to the orbital motion, as well as
circularized the orbit despite the wider separation. Whether the
formal eccentricity is real or simply related to surface activity
should be checked with further observations. 

 \noindent{\bf [22]~RXJ0408.6$+$1017.~} Single-lined binary similar to
[20]~RXJ0403.5$+$0837, and with nearly the same period. The minimum
radius implied by the measured $v \sin i$ is $R_A \sin i =
1.42$~R$_{\sun}$, which is larger than expected for a star with a
temperature slightly cooler the Sun.  Some evolution is likely, and a
value of $\log g = 4.0$ is adopted. 

 \noindent{\bf [23]~RXJ0410.6$+$0608.~} The measured projected
rotational velocity of the primary is small but rather uncertain in
this SB1. The temperature corresponds to a spectral type near K2 for a
main-sequence star. Adopting a typical mass of $M_A \sim
0.76$~M$_{\sun}$, the secondary star would be of spectral type M3 or
earlier. The possibility suggested by \markcite{m97}M97 that this is a
PMS system, based on their detection of weak \ion{Li}{1}~$\lambda$6708
absorption, seems unlikely. 

 \noindent{\bf [24]~RXJ0422.9$+$0141.~} As in [16]~RXJ0343.6$+$1039,
the mass ratio in this double-lined system is close to unity but the
secondary is much brighter than the primary ($L_B/L_A = 4.1$ at a mean
wavelength of 5187~\AA), and is also hotter. The primary must be
evolved, perhaps being an early K giant or subgiant. \markcite{m97}M97
reported weak \ion{Li}{1}~$\lambda$6708 absorption in both components,
and advanced the possibility that it might be a PMS system.
\markcite{n97}N97, on the other hand, considered it to be a zero-age
main sequence (ZAMS) system with an age of the order of 30~Myr, but
based only on the temperature of the hotter secondary.  From the
properties discussed above, the PMS scenario can be ruled out, and a
post-main sequence picture seems more likely. Further support for this
is given in \S\ref{secdensity} by comparison with evolutionary tracks. 

Using eq.(3) the difference in surface gravities is $\Delta\log g
\approx 0.8$, the primary value being lower. The system is detached,
and $i_{\rm min}$ is about 70\arcdeg. A normal main sequence mass for
the secondary of 1.3~M$_{\sun}$ corresponding to its measured
temperature (mid F star) would imply an inclination angle close to
63\arcdeg, which would make eclipses unlikely. With this inclination
the sizes of the stars would be $R_A \sim 4.4$~R$_{\sun}$ and $R_B
\sim 1.8$~R$_{\sun}$, and the primary mass would be 1.35~M$_{\sun}$.
The larger radius of the secondary compared to a normal F star would
indicate that it too is somewhat evolved. The unusually large scatter
of the primary velocities from the orbital fit may be a sign of
surface activity in the form of spots. Overall this system is
remarkably similar to [16]~RXJ0343.6$+$1039, except perhaps for the
higher inclination angle. 

 \noindent{\bf [25]~RXJ0427.8$+$0049.~} Synchronization in this short
period SB2 system with a circular orbit leads to minimum radii of $R_A
\sin i = 2.0$~R$_{\sun}$ and $R_B \sin i = 1.4$~R$_{\sun}$. The value
for the secondary is only slightly larger than normal for a late F or
early G main-sequence star, but perhaps consistent if one accounts for
the measurement errors in $v \sin i$. The primary, on the other hand,
appears considerably larger than normal, and is likely to be evolved.
An assumed mass of 1.1~M$_{\sun}$ for the secondary based on its
effective temperature leads to an inclination angle of about
69\arcdeg, whereas the minimum angle for eclipses to occur is
75\arcdeg. The system is well detached, and the light ratio is
$L_B/L_A = 0.67$. Alternate designations are BD$+$00$\arcdeg$760 and
TYC~75-1529-1. 

 \noindent{\bf [26]~RXJ0429.9$+$0155.~} Both stars in this long-period
double-lined binary are slow rotators. The minimum masses seem large
for main-sequence stars of the temperatures that we determine (SpT
G9-K0). The stars are possibly evolved. The mass ratio is close to
unity ($q\equiv M_B/M_A = 0.94$, and the light ratio is $L_B/L_A =
0.24$. The system is well detached, and is also known under its Tycho
designation TYC~75-1-1. 

 \noindent{\bf [27]~RXJ0434.3$+$0226.~} The temperature we determine
for the primary in this very long period, eccentric single-lined
binary corresponds to spectral type K2, if it is a main sequence star.
In that case, the minimum mass of the secondary corresponds to
spectral type M3.  Little more can be said about the nature of the
system from the dynamical information alone. \markcite{m97}M97
measured moderate \ion{Li}{1}~$\lambda$6708 absorption and considered
the object to be in the PMS phase. Our own measurements confirm the
presence of Li, which is stronger than the nearby
\ion{Ca}{1}~$\lambda$6718 line. The classification was refined by
\markcite{m99}Magazz\`u, Umana \& Mart\'\i n (1999), who listed it as
a post-T~Tauri star (PMS star with significant Li depletion).
\markcite{n97}N97 reported a similar conclusion, estimating the age to
be less than $\sim$30~Myr.  Consequently, the surface gravity we have
adopted ($\log g = 4.0$) is slightly lower than that corresponding to
a normal main-sequence star. 

 \noindent{\bf [28]~RXJ0435.5$+$0455.~} Similar to previous system,
with a K0 primary (if on the main sequence) and a minimum secondary
mass corresponding roughly to an M4 star. Tycho designation
TYC~90-936-1. 

 \noindent{\bf [29]~RXJ0441.1$+$1132.~} The secondary in this
short-period SB2 is very faint ($L_B/L_A = 0.07$), and its temperature
and projected rotational velocity are very uncertain. The assumption
of synchronization and a normal radius for a main-sequence star with
the temperature we determine for the primary lead to a rotational
velocity for that star that is slightly lower than actually measured,
but possibly consistent within the errors. The minimum mass for the
primary from our spectroscopic orbit is also close to the typical
value for that temperature, so that the data are consistent with a K0
primary in an orbit viewed nearly edge-on. The system is detached.
However, we cannot rule out lower inclinations that would give larger
values of the mass and radius, implying some measure of evolution. 

The eccentricity we determine is only marginally significant ($e =
0.0126 \pm 0.0049$, 2.6$\sigma$), and is possibly an artifact from
distortions in the velocities related to surface activity, given the
short orbital period (2.55~days) and convective envelopes of the
stars. Alternate designation TYC~690-1116-1.

 \noindent{\bf [30]~RXJ0441.9$+$0537.~} This double-lined binary (also
BD$+$05$\arcdeg$706) is an Algol-type system (semi-detached) of a rare
class referred to as ``cool Algols", of which only about a dozen
examples are known.  Both components are late-type giants, as opposed
to one of them being an early-type dwarf, as in the ``classical"
Algols, and the system has undergone (and is perhaps still undergoing)
mass transfer.  Full details were reported by \markcite{t98}Torres,
Neuh\"auser \& Wichmann (1998). 

 \noindent{\bf [31]~RXJ0442.3$+$0118.~} SB2 system (TYC~83-788-1) with
a circular orbit and a rather low mass ratio, indicating that one or
both components are evolved. The light ratio is also quite small
($L_B/L_A = 0.16$).  The cooler primary temperature (spectral type
early K) suggests that this star must be the larger one, consistent
with its larger mass. Its measured $v \sin i$ inserted into eq.(3)
supports this, giving a $\log g$ value under 3.5, which we have
adopted. The system is well detached.  If the secondary is a main
sequence star, a normal mass of $M_B \sim 0.81$~M$_{\sun}$ for its
temperature suggests an inclination angle near $73\arcdeg$, although
it could be larger. The minimum inclination angle for eclipses to
occur, from eq.(3), is $i_{\rm min} \approx 77\arcdeg$. The
possibility of eclipses cannot be ruled out, and photometric
observations are underway to investigate this. 

 \noindent{\bf [32]~RXJ0442.6$+$1018.~} The rotation of the primary in
this long-period SB1 is small, but uncertain. Under the assumption
that it is a main-sequence star (SpT $\sim$ K2), the minimum mass for
the secondary corresponds to a late-type M star. Tycho designation
TYC~686-1246-1. 

 \noindent{\bf [33]~RXJ0442.9$+$0400.~} Another object presenting
triple-lined spectra. The velocities for the main component give a
nearly circular orbit with a period of 510 days. The spectral type is
approximately G7. The secondary component appears to be itself a
double-lined binary, but the period is unknown. Weak
\ion{Li}{1}~$\lambda$6708 absorption was detected by
\markcite{m97}M97, who suspected a possible PMS nature, while
\markcite{n97}N97 considered it to be a $\sim$100~Myr-old ZAMS system.
The entry in the Tycho catalogue for this object (TYC~91-702-1)
suggests photometric variability, as indicated by the scatter of
$\sim$0.9~mag in the $V_T$ passband.  Follow-up is warranted.

 \noindent{\bf [34]~RXJ0444.3$+$0941.~} The lines of the primary star
in this single-lined system are very broad, consistent with the short
period of the binary. The temperature we determine corresponds
approximately to spectral type F7. Synchronization and a typical
radius for such a star (1.22~R$_{\sun}$) lead to an equatorial
rotational velocity of 90\kms. Our measured $v \sin i = 120\kms$ then
implies an inclination angle near 49\arcdeg. If the primary mass is
normal (1.27~M$_{\sun}$), the secondary is of spectral type M3 or
earlier, and is likely to be also a fast rotator.  This system has
also received the designation HD~287017. 

 \noindent{\bf [35]~RXJ0444.4$+$0725.~} Short-period SB2 with a
circular orbit. The components' rotation is most likely synchronized
with the orbital motion. The secondary star is rather faint ($L_B/L_A
= 0.13$) and its parameters correspondingly more uncertain. The mass
ratio and light ratio are consistent with the standard mass-luminosity
relation for dwarfs.  Eq.(3) and the measured rotation of the primary
also lead to a normal surface gravity for a main-sequence star. A
typical mass for the primary of $M_A \sim 0.75$~M$_{\sun}$, based on
its temperature, suggests an inclination angle of $i \sim
68\arcdeg$. However, the minimum angle for eclipses to occur in this
system is $i_{\rm min} \approx 78\arcdeg$, so that eclipses seem to be
ruled out. The system is well detached. Only weak
\ion{Li}{1}~$\lambda$6708 absorption was detected by
\markcite{m97}M97, so that a PMS status for the object seems unlikely
(see also \S\ref{secdensity}).

 \noindent{\bf [36]~RXJ0451.6$+$0619.~} Single-lined binary with a
circular orbit and a very short period.  The temperature of the
rapidly rotating primary corresponds to a spectral type of G8.
Synchronization and a typical radius for a main-sequence star
(0.88~R$_{\sun}$) gives an equatorial rotational velocity of about
64\kms, whereas the value we measure is $v \sin i = 81$\kms. The star
may be slightly evolved (larger). There is no clear sign of the
secondary star in our spectra, despite the large value we infer for
its minimum mass ($M_B \sin i \approx 0.76$~M$_{\sun}$), assuming a
normal mass for the primary. A faint companion approximately
$1\arcsec$ in the ENE direction was seen at the telescope. 

 \noindent{\bf [37]~RXJ0503.6$+$1259.~} The secondary in this SB2 is
extremely faint ($L_B/L_A = 0.07$), and its temperature and rotational
velocity are very uncertain. From the very small mass ratio at least
one of the stars must be evolved. Despite the short period the orbit
appears to be slightly eccentric, although further observations are
needed to confirm this. If real, the spins of the stars are likely to
be synchronized with the orbital motion at periastron. In that case
the measured $v \sin i$ of the primary translates into a size that is
very close to that of the critical Roche surface for that star.
Because the minimum masses are so small, the inclination angle is also
likely to be small ($10\arcdeg$ to $30\arcdeg$ for any reasonable
values for the absolute masses).  Therefore, the contribution of the
last term in eq.(3) is not negligible ($\approx$0.5~dex). Accounting
for this, we have adopted $\log g_A = 3.0$. The nature of the
secondary is unknown.  The system appears to be similar to the Algols,
or is possibly a W~UMa-type object. 

 \noindent{\bf [38]~RXJ0515.3$+$1221.~} This object (also known as
TYC~707-1311-1) has a complicated spectrum that appears to be
triple-lined. The primary velocities yield an orbit with a period of
63 days. The secondary in this system is itself a binary, apparently
double-lined.  Although it has not been possible to determine its
orbit, the period must be short since the velocity amplitudes of the
secondary and tertiary components are at least 70\kms. This is likely
to be the source of the X-ray radiation detected by ROSAT. There is a
faint visual companion approximately $20\arcsec$ the south of this
triple system. 

 \noindent{\bf [39]~RXJ0523.0$+$0934.~} The short period and circular
orbit make synchronization very likely in this SB1 system.  The
measured $v \sin i$ is in good agreement with the value predicted
assuming a normal radius of 1.25~R$_{\sun}$ for a main-sequence star
of the temperature we measure (which corresponds to \ion{F6}{5}, and
1.3~M$_{\sun}$). This, in turn, implies an orbit viewed nearly
edge-on. The minimum mass for the secondary corresponds roughly to an
M3 star. The object (TYC~704-975-1) is listed in the Tycho catalogue
as having a scatter of 0.4~mag in the $V_T$ passband, although this
needs confirmation. The system is most likely detached. 

 \noindent{\bf [40]~RXJ0523.5$+$1005.~} The rapid rotation of the
components in this SB2 system is due to the very short orbital period
(0.55~days).  However, the predicted equatorial rate assuming
synchronization is nearly twice the measured $v \sin i$ values,
implying a fairly low inclination angle.  Adopting masses for
main-sequence stars (0.9~M$_{\sun}$ and 0.8~M$_{\sun}$) consistent
with the measured temperatures, we derive $i \sim 27$\arcdeg, although
this inclination together with our measured $v \sin i$ values gives
radii slightly larger than normal for stars of this spectral type (G7
and K0): 1.2~R$_{\sun}$ and 0.9~R$_{\sun}$.  Eclipses are not expected
for inclinations lower than $i_{\rm min} \approx 50$\arcdeg.  The
system is probably still detached, although the primary star appears
to be rather close to filling its Roche lobe (90\%). The light ratio
is $L_B/L_A = 0.22$. Tycho designation TYC~704-2521-1. 

 \noindent{\bf [41]~RXJ0528.9$+$1046.~} The assumption of
synchronization in this 7.7-day period double-lined binary along with
our measured values of $v \sin i$ implies low surface gravities for
both components. From eq.(3) we derive $\log g = 3.5$, ignoring the
contribution from the term that depends on the unknown inclination.
The system is well detached, and the light ratio at 5187~\AA\ is
$L_B/L_A = 0.34$.  The lower limit for eclipses to occur is $i_{\rm
min} \approx 77\arcdeg$, but the small minimum masses suggest that the
actual inclination is considerably lower than this.  Our formal
spectroscopic solution gives an orbit with a small but apparently
significant eccentricity: $e = 0.0290 \pm 0.0071$ (4$\sigma$). This is
somewhat unexpected for the relatively short period and the fairly
deep convective envelopes of the late-type components, and could be
caused by distortions in the measured radial velocities due to
chromospheric activity (spots). Even if the eccentricity is real, the
spins of the stars are likely to be synchronized with the orbital
motion at periastron, and the values inferred above for the surface
gravity still hold. 

From the dynamical evidence the stars are larger than normal dwarfs.
One possibility is that they are both subgiants or giants of spectral
type early- or mid-K, quite typical of the RS~CVn class. However, the
presence of very strong \ion{Li}{1}~$\lambda$6708 absorption and other
clues suggest otherwise. The object is projected against the
$\lambda$~Ori star-forming region, and its center-of-mass velocity of
$+28.3 \pm 0.2$\kms\ is consistent with the mean velocity of other
objects in that complex.  \markcite{m97}M97 measured a Li absorption
strength of 0.4~\AA, at an epoch when the spectral lines of the two
components were severely blended.  More recently \markcite{dm01}Dolan
\& Mathieu (2001) established that both stars show very prominent Li
lines. After correction for binarity using an estimated light ratio of
$L_B/L_A = 0.5$ at $\lambda$6708, their raw measurements correspond to
equivalent widths 0.40~\AA\ and 0.45~\AA\ for the primary and
secondary, respectively.  The Li excess over the levels displayed by
stars of similar temperatures in young clusters such as the Pleiades
or IC~2602 is strongly indicative of a very young age (see
\markcite{n97}N97), and supports its classification as a wTTS (see
also \markcite{m99}Magazz\`u et al.\ 1999).  $V\!RI$ photometry by
\markcite{dm01}Dolan \& Mathieu (2001), along with an assumed distance
of 450~pc for the $\lambda$~Ori region, led them to an estimated age
of 2.3~Myr for RXJ0528.9$+$1046, using the stellar evolution models by
\markcite{ps99}Palla \& Stahler (1999).  The conclusion of the PMS
nature for this object is further supported by our independent
comparison with evolutionary tracks in \S\ref{secdensity}. 

 \noindent{\bf [42]~RXJ0529.3$+$1210.~} The orbit in this long-period
binary is extremely eccentric ($e \geq 0.9$), and because of this the
elements are still preliminary. The appearance of the correlation
functions and the large minimum secondary mass we derive from our
single-lined orbit suggest there may be light from another star, or
possibly from two additional stars. The temperature of the primary
corresponds roughly to a spectral type of K6, if it is a dwarf. The $v
\sin i$ value we measure ($18\kms$), however, is higher than expected
for a normal main-sequence star.  It is possible that the axial
rotation is synchronized with the orbital motion at periastron
(``pseudo-synchronized"), which, by virtue of the exceptionally large
eccentricity, would be considerably faster than the mean orbital
motion by a factor of more than 50. If that is the case, the projected
radius of the star would be $R \sin i = 2.9$~R$_{\sun}$. This is much
larger than normal for a K6 dwarf, suggesting in principle either a
post-main sequence or a pre-main sequence status.

As in the previous system, RXJ0529.3$+$1210 is projected against the
$\lambda$~Ori complex, in this case towards the edge and coinciding
with a density enhancement in the CO maps for this region (see
\markcite{dm01}Dolan \& Mathieu 2001). The center-of-mass velocity of
$+18.7 \pm 0.7$\kms, however, is more consistent with the mean for the
Taurus-Auriga region, as noticed also for other stars in this area by
\markcite{dm01}Dolan \& Mathieu (2001).  This circumstantial evidence
favors the interpretation that this is a young system, rather than a
highly evolved (post-main sequence) object.  The Lithium strength as
measured by \markcite{m97}M97 is 0.35~\AA, while we measure a slightly
lower value of 0.27~\AA. Given the cool temperature of the primary,
this further supports the notion that it is a relatively young system
since cooler objects tend to deplete their primordial Lithium quite
rapidly because of their deeper convective envelopes (see
\S\ref{secpms}), and would not be expected to show such high levels.
The H$\alpha$ emission is among the strongest in our sample. Both
\markcite{m97}M97 and \markcite{n97}N97 considered it to be a PMS
object.  \markcite{m99}Magazz\`u et al.\ (1999) listed it as a
post-T~Tauri star. 
	
 \noindent{\bf [43]~RXJ0530.9$+$1227.~} SB1 with a short period and a
circular orbit. If the rotation of the primary is locked with the
orbital motion, as expected, the lower limit to the radius based on
eq.(2) and our measured $v \sin i$ is $R_A \sin i = 1.8$~R$_{\sun}$.
This is about a factor of 2 larger than normal for a main sequence
star with the temperature we determine, which indicates that the star
is probably evolved (spectral type late G). Accordingly, we have
adopted a lower surface gravity of $\log g = 4.0$. The residuals from
the orbit show a pattern suggestive of a long-term variation ($P >
5$~yr). The center-of-mass velocity of this system, $+92.4\kms$, is
the largest in our sample. 

\section{Discussion}

\subsection{Lithium strengths and PMS status}
\label{secpms}

The original selection of the \markcite{n97}N97 sample, while designed
to favor the detection of young stars in a flux-limited X-ray survey
(RASS), has obviously not been completely successful as attested by
the large fraction of (post-main sequence) RS~CVn systems among our
binaries.  Nevertheless, a handful of our program stars do have
detectable \ion{Li}{1}~$\lambda$6708 absorption, one of the key
diagnostics of youth used to support PMS status. Because the strength
of the Lithium line depends quite strongly on the effective
temperature, and even on the rotational velocity of the star (see,
e.g., \markcite{s93}Soderblom et al.\ 1993; \markcite{mc96}Mart{\'\i}n
\& Claret 1996), both of these parameters must be accounted for to
evaluate the significance of the Li signature.  Even with this,
additional information on the systems can also be very helpful, such
as that provided by our orbital solutions in \S\ref{secresults} and
\S\ref{seccomments}. 

\begin{figure}
 \plotfiddle{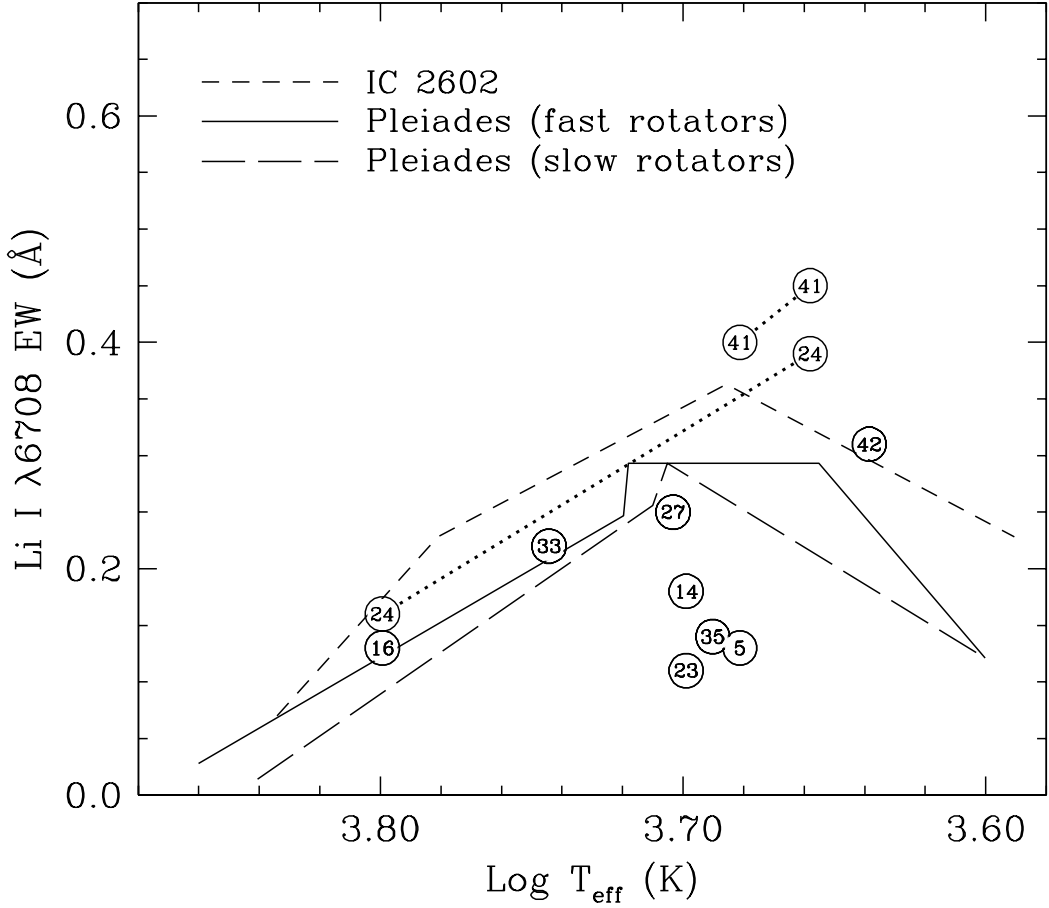}{5.0in}{0}{85}{85}{-260}{-160}
\figcaption[Torres.fig05.ps]{\ion{Li}{1}~$\lambda$6708 equivalent widths from
Table~\ref{tabsample} for all systems in our sample with
determinations exceeding 0.1~\AA\ (average values are taken when more
than one measurement is available).  The segmented lines represent the
upper envelopes of the Li measurements as a function of temperature
for the young cluster IC~2602 ($\sim$35~Myr) and for the Pleiades
($\sim$100~Myr) following N97. For the latter cluster separate
envelopes are indicated for the rapid rotators and the slow rotators,
with the dividing line set at $v \sin i = 15$\kms. The systems are
identified by the running number in Table~\ref{tabsample}.  Dotted
lines connect the components of the double-lined binaries where
separate measurements exist for the two stars.\label{figli}}
 \end{figure}

In Figure~\ref{figli} we represent the Li measurements as a function
of effective temperature for the 10 objects in our sample with
equivalent widths larger than 0.1~\AA, as listed in
Table~\ref{tabsample}. For stars with more than one measurement we
have adopted the average. Six systems are double-lined, but only two
have separate measurements for the two components. For the others, we
have adopted the temperature of the brighter component and assumed
that the Li measurement corresponds largely to that star. The
measurements for all double-lined systems have been corrected for the
dilution factor due to binarity, using our temperature determinations
and approximate light ratios at 6708~\AA\ estimated from our
measurements at 5187~\AA. 

\placefigure{figli}

The segmented lines in the Figure~\ref{figli} represent the upper
envelope of the Li distributions for two young clusters often used for
comparison purposes --- IC~2602 (age $\sim 35$~Myr), and the Pleiades
(age $\sim 100$~Myr) --- following \markcite{n97}N97.  For the latter
cluster, which has been studied in considerably more detail, the rapid
rotators ($v \sin i > 15\kms$) display somewhat lower Li depletion
(larger equivalent widths) than the slow rotators, as represented by
the solid and long-dashed lines, respectively. 

Three of our systems deserve special comment. The cooler component of
[24]~RXJ0422.9$+$0141 lies slightly above the upper envelope for
IC~2602, suggesting the possibility of a PMS status. However, the
dynamical and physical information discussed in \S\ref{seccomments}
seems to rule this out (see also below), since the larger mass for
this cooler component indicates that this star must have evolved to
become a subgiant or a giant. We conclude therefore that this is a
post-main sequence system of the RS~CVn type. Studies by
\markcite{prg92}Pallavicini, Randich \& Giampapa (1992),
\markcite{ff93}Fern\'andez-Figueroa et al.\ (1993),
\markcite{r94}Randich, Giampapa \& Pallavicini (1994) and others have
shown that active systems of the RS~CVn type and other categories of
active objects occasionally show unexpectedly high levels of Li
absorption compared to inactive stars of similar spectral types. While
the mechanism responsible for the excess Li abundance is not yet
clear, the case of [24]~RXJ0422.9$+$0141 appears consistent with those
findings. 

The two components of [41]~RXJ0528.9$+$1046 show significant excess Li
absorption, and the supplementary evidence discussed in
\S\ref{seccomments} strongly supports its classification as a
weak-lined T~Tauri system.  The Li strength measured for
[42]~RXJ0529.3$+$1210 puts it essentially at the upper envelope of
IC~2602, but this may be only a lower limit since there are hints in
our spectra of light from one or perhaps two additional components.
This extra light would cause the true equivalent width of the primary
to be underestimated, though it is difficult at the moment to
determine by how much. Based on this and the strong suggestion of
association with the $\lambda$~Ori region discussed in
\S\ref{seccomments}, we consider it quite likely that this system is
also very young. 

Thus, two of our program stars display properties indicating that they
are PMS objects with ages most likely under $\sim$30~Myr. The fact
that they are binaries makes them especially interesting, given that
only about three dozen such objects have had their spectroscopic
orbital elements determined. Furthermore, [42]~RXJ0529.3$+$1210 is
among only a handful with periods long enough that the components may
be spatially resolved in the near future using large ground-based
interferometers observing in the infrared.  At these wavelengths the
contrast between the primary and secondary components should be much
more favorable than in the optical. At an assumed distance of 450~pc
for [42]~RXJ0529.3$+$1210 (\markcite{dm01}Dolan \& Mathieu 2001) the
angular semimajor axis of the pair is estimated to be about 3~mas. But
because of the large eccentricity of its orbit, the maximum separation
can be as large as twice this value. 
		
\subsection{A comparison with evolutionary tracks}
\label{secdensity}

In \S\ref{seccomments} we made use of the measured $v \sin i$ values
for the short-period double-lined binaries in our sample that have
circular orbits, in order to provide estimates of the radii and surface
gravities of the components and help in understanding their nature.
In a sense the rotational velocities have thus been used as a
sensitive measure of evolution, under the reasonable hypothesis that
tidal forces have already synchronized the rotation of the stars and
aligned their axes with that of the orbit. Those estimates are only
lower limits, though, because of the unknown inclination angles. In
this section we take this approach one step further and eliminate the
dependence on $i$ to provide a physical magnitude that may be compared
directly with predictions from theory for normal stars. 

Eq.(1) and eq.(2) give the quantities $M \sin^3 i$ and $R \sin i$
directly in terms of observable properties ($P$, $K$, and $v \sin i$).
The minimum masses and minimum radii have different dependences on
$\sin i$, but the ratio $M \sin^3 i/(R \sin i)^3$ removes the
dependency, and happens to represent the \emph{density} of a star
($M/R^3$) in terms of the solar density.  Expressed directly as a
function of the measurable parameters, we have
 \begin{equation}
\log {\rho_{A,B}\over\rho_{\sun}} = -1.8718 - 2\log P + 2\log(K_A+K_B) + \log K_{B,A} - 3\log V^{\rm rot}_{A,B}~,
\end{equation}
 where all quantities are given in their usual units. The density is
intimately related to the internal structure and evolutionary status
of a star, and the possibility of using this ``observable" property in
suitable double-lined systems to compare directly with stellar
evolution models is often overlooked\footnote{For a recent application
of this idea, see \markcite{q00}Quast et al.\ (2000).}.  The density
changes by more than two orders of magnitude over the life of a normal
star, so that at the very least this allows us (in conjunction with
the measured effective temperatures) to tell whether the components
are still on the main sequence or whether they are evolved. As it
turns out, the discriminating power of the density is quite
significant although it depends greatly on the precision of the
rotational velocities, as seen from the factor of 3 in the last term
of eq.(6). 

\begin{figure}
 \plotfiddle{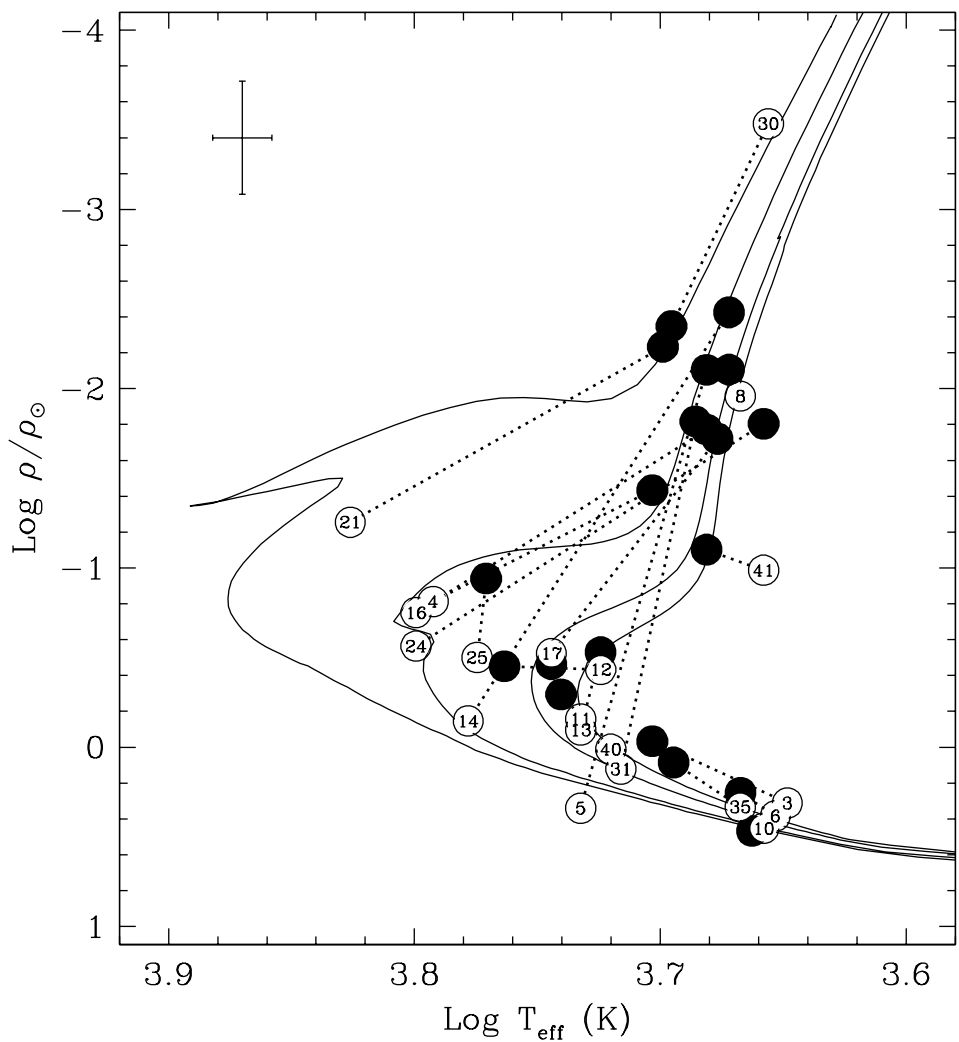}{5.5in}{0}{75}{75}{-240}{-100}
\figcaption[Torres.fig06.ps]{Double-lined binaries in the $\log\rho/\rho_{\sun}$
vs.\ $\log T_{\rm eff}$ plane, compared to theoretical isochrones for
the main sequence and post-main sequence phase by Yi et al.\ (2001)
for solar metallicity and ages of 1~Gyr, 3~Gyr, 10~Gyr, and 15~Gyr
(from left to right). The primary components (filled symbols) are
connected to the secondaries (open symbols) by a dotted line. Systems
are identified by the running number in Table~\ref{tabsample}. Average
errors are indicated on the upper left, but vary significantly from
system to system.\label{figdens}}
 \end{figure}

In Figure~\ref{figdens} we show the components of all the double-lined
systems in our sample that have sufficiently well determined
rotational velocities, in a diagram of $\log \rho/\rho_{\sun}$ vs.\
$\log T_{\rm eff}$. In all cases the periods are short or there is
otherwise good reason to believe that the assumption of
synchronization and alignment is valid, as described in the individual
notes for each system. In each binary a dotted line connects the
primary components (dark symbols) with the secondaries (open symbols,
with the identification number).  The observations are compared with
theoretical isochrones for the main-sequence and post-main-sequence
phase based on the models by \markcite{y01}Yi et al.\ (2001), for
solar metallicity and ages of 1~Gyr, 3~Gyr, 10~Gyr and 15~Gyr. 

\placefigure{figdens}

Almost all of the secondary components are seen to lie near the main
sequence, while roughly half of the primaries are evolved and lie at
the base of the giant branch or above. A gap is seen in the diagram
between the evolved and un-evolved stars that coincides with the locus
of objects in the rapid subgiant phase, where fewer stars are expected
at any given time. For the most part the two components of each binary
are consistent with being on a single isochrone, given the errors, the
most obvious exceptions being [12]~RXJ0309.1$+$0324N and
[41]~RXJ0528.9$+$1046, which seem to be aligned in a direction roughly
perpendicular to the isochrones. The first of these is most likely a
W~UMa-type system, and its components should therefore not be expected
to behave as single stars since they are in close mechanical and
thermal contact.  The evidence presented earlier for the other system
indicates that it is likely to be in the PMS stage (see below).  But
for all the other objects, we note that the location of each star is
fully consistent with our interpretation in \S\ref{seccomments}.  This
is expected since the same hypothesis was used, but the evolutionary
status of each binary is more straightforward to see and to compare
with the models in this figure. 

The densities of the primary components in all the binaries are
typically smaller than those of the secondaries, as dictated by
stellar evolution, except in the case of [30]~RXJ0441.9$+$0537, which
is the highest object on the diagram in Figure~\ref{figdens}.  In this
system (which is the ``cool Algol" referred to earlier) mass transfer
has caused a reversal of the mass ratio, so that the currently more
massive star is actually the one that was originally the secondary
(see \markcite{t98}Torres, Neuh\"auser \& Wichmann 1998). 

\begin{figure}
 \plotfiddle{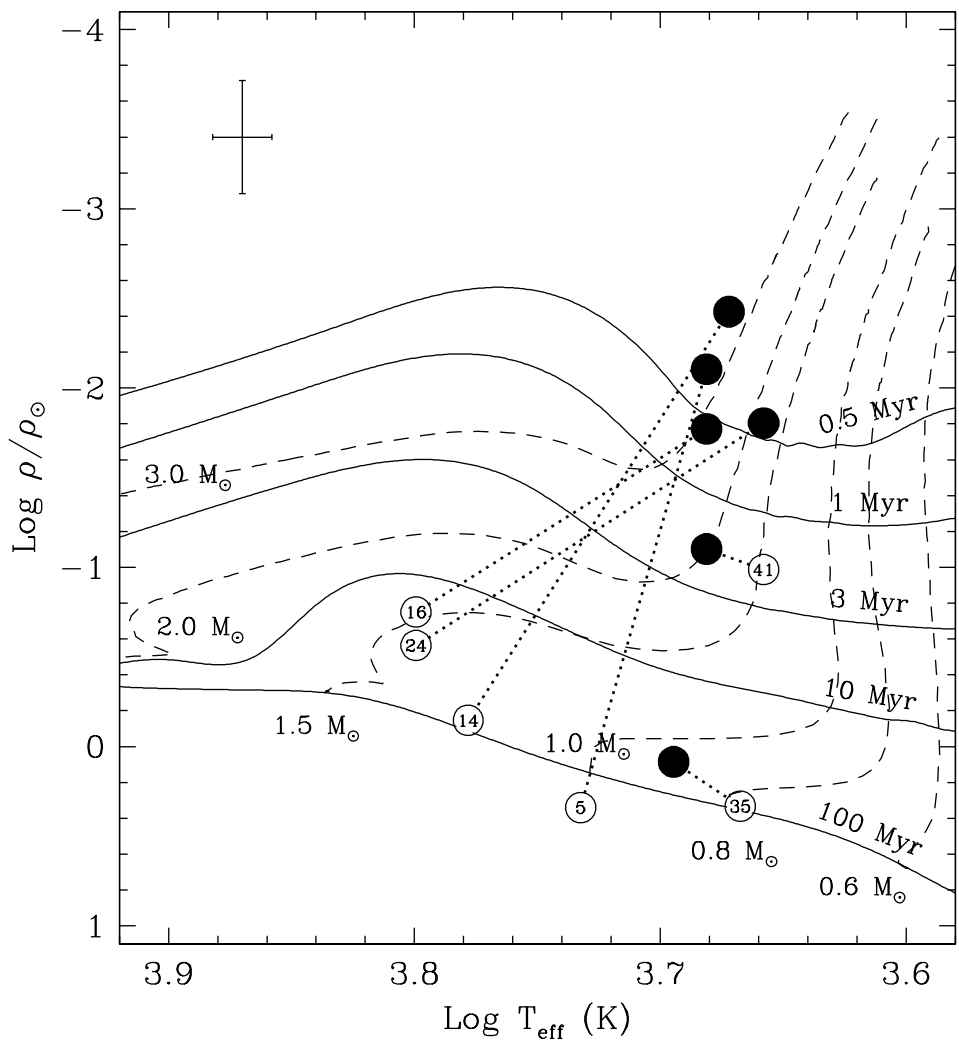}{5.5in}{0}{75}{75}{-240}{-100}
\figcaption[Torres.fig07.ps]{Same as Figure~\ref{figdens}, but for PMS
evolutionary tracks and isochrones from Siess, Forestini \& Dougados
(1997) for solar metallicity. Isochrones (solid curves) and mass
tracks (dashed) are as labeled. All double-lined systems in our sample
with measured \ion{Li}{1}~$\lambda$6708 absorption larger than
0.1~\AA\ are represented for comparison (see text).\label{figdenspms}}
 \end{figure}

\placefigure{figdenspms}

As mentioned earlier, a few of our systems exhibit moderately strong
\ion{Li}{1}~$\lambda$6708 absorption that may indicate an early stage
of evolution (PMS), as opposed to a post-main sequence status. Six of
them are double lined. It is quite illuminating to represent these
systems in a $\log\rho/\rho_{\sun}$ vs.\ $\log T_{\rm eff}$ diagram
such as Figure~\ref{figdens}, but to compare them instead with
evolutionary tracks appropriate for the PMS stage. This is done in
Figure~\ref{figdenspms}, where the isochrones and evolutionary tracks
are from the models by \markcite{s97}Siess, Forestini \& Dougados
(1997) (see also \markcite{s00}Siess, Dufour \& Forestini 2000) for
solar metallicity. 

The components of [5]~RXJ0219.7$-$1026, [14]~RXJ0339.6$+$0624,
[16]~RXJ0343.6$+$1039, and [24]~RXJ0422.9$+$0141 are clearly
inconsistent with being on the same isochrone, and this agrees with
our assessment in \S\ref{seccomments} that they are not likely to be
in the PMS phase based on dynamical and physical evidence (e.g., the
mass ratio and luminosity ratio), or based on their relatively weak Li
line in most cases (\S\ref{secpms}).  Objects [35]~RXJ0444.4$+$0725
and [41]~RXJ0528.9$+$1046, on the other hand, do seem to conform to
the isochrones in this diagram, within the errors. The age inferred
for the first of these is not particularly young, and its properties
(including the weak Li absorption) are consistent with it being a ZAMS
star.  The location in the $\log\rho/\rho_{\sun}$ vs.\ $\log T_{\rm
eff}$ plane is also consistent with the main-sequence isochrones shown
in Figure~\ref{figdens}, so that the case for a PMS status is not very
strong in this system. 

More interesting is the agreement of [41]~RXJ0528.9$+$1046 with the
PMS isochrones in Figure~\ref{figdenspms}, while running \emph{across}
those in Figure~\ref{figdens}.  This, along with the Li strengths and
kinematic information reported earlier, makes a more compelling case
for this system being a much younger object.  The absence of H$\alpha$
emission reported by \markcite{m97}M97 and by \markcite{dm01}Dolan \&
Mathieu (2001) place it in the wTTS category.  The age we derive for
this SB2 from Figure~\ref{figdenspms} is approximately 2~Myr,
consistent with the estimate of 2.3~Myr by \markcite{dm01}Dolan \&
Mathieu (2001) based on a different set of models
(\markcite{ps99}Palla \& Stahler 1999). Use of yet another set of PMS
isochrones by \markcite{y01}Yi et al.\ (2001) yields a younger age of
about 1~Myr. The absolute masses for the components from this figure
($M_A = 1.96$~M$_{\sun}$ and $M_A = 1.46$~M$_{\sun}$) are not quite
consistent with the spectroscopic mass ratio, which is measured much
more accurately. This is most likely due to errors in the temperature
determinations. In addition, the absolute masses are strongly
model-dependent. For example, the values inferred from the PMS
evolutionary tracks by \markcite{y01}Yi et al.\ (2001) are
systematically lower by 0.4-0.5~M$_{\sun}$, which is rather
significant. Nevertheless, one may derive an average inclination angle
for the orbit that is roughly 33-37\arcdeg, which rules out eclipses,
as indicated in \S\ref{seccomments}. 

It is worth pointing out, to conclude this section, that a diagram
such as Figure~\ref{figdenspms} can be a very useful tool in the field
of young stars, specifically for the case of double-lined
spectroscopic binaries with short periods and circular orbits. It is
similar to the more widely used H-R diagram, but has the important
advantage that it is completely independent of any assumption on the
distance, a common criticism that is painfully familiar to all users
of the $\log L/L_{\sun}$ vs.\ $\log T_{\rm eff}$ diagram. 
	
\subsection{Period distribution and binary frequency}
\label{secpdistrib}

In the course of solving for the orbits of the binaries in the
\markcite{n97}N97 sample we had the distinct impression that the
fraction of short-period systems was unusually high. Due to the nature
of the target list, which was selected to include objects that are
strong in X-rays, a bias towards short periods is actually expected
from the fact that those systems will tend to be rotating faster
because they are tidally locked, and thus will have enhanced activity
and should be detected more easily by ROSAT.  As it turns out, more
than half of our binaries have orbital periods of less than 5 days. In
addition, we noticed two other trends that were not expected for field
stars, but in retrospect can also be understood in terms of selection
effects.  One is that the frequency of binaries among the ROSAT
sources south of Taurus seemed higher than normal (43 spectroscopic
binaries with orbits, out of a sample of 121 stars), and the other is
that the fraction of double-lined systems (as opposed to single-lined
binaries) also seemed high: 26 SB2 systems out of 43 binaries, or
60\%. 

The detection of spectroscopic binaries in our sample is by no means
complete as a function of orbital period. The majority of the stars
were observed over an interval of 3 to 5 years, and we estimate
conservatively that we have detected most of the systems with orbital
periods up to about 1~yr, although we also found 3 binaries with even
longer periods. The incompleteness due to small velocity amplitudes
(low-mass companions, small inclinations, or both) is difficult to
quantify, but it is unlikely to be very large for this sample and will
not be considered here. 

\begin{figure}
 \plotfiddle{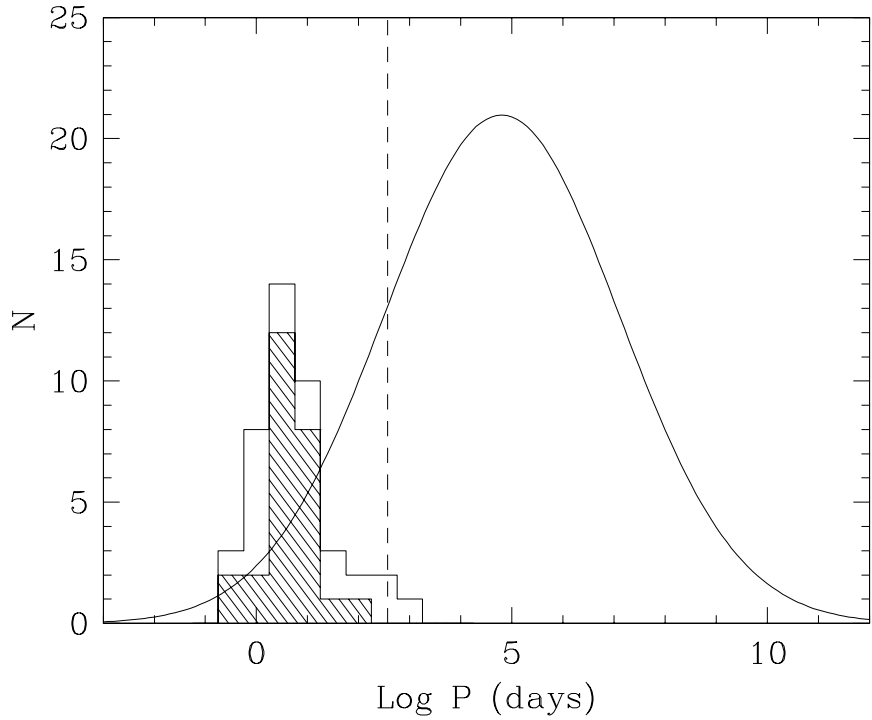}{4in}{0}{75}{75}{-240}{-170}
\figcaption[Torres.fig08.ps]{Period distribution (on a logarithmic scale) of
the 43 binaries in the present sample compared to the solar-type
binaries in the solar neighborhood. The hatched area of the histogram
corresponds to the double-lined binaries in our sample, and the open
area to the single-lined systems. The Gaussian curve representing the
field binaries is taken from Duquennoy \& Mayor (1991), and was
normalized to yield the same number of systems that we detect up to
our completness level of 1~yr, indicated by the dashed line. The
excess of short-period systems in the X-ray sample is discussed in the
text.\label{figpdist}}
 \end{figure}

As a benchmark for studying the frequency and period distribution of
the binaries in our X-ray sample we have used the results for the
solar-type stars in the solar neighborhood by \markcite{dm91}Duquennoy
\& Mayor (1991). In Figure~\ref{figpdist} we show the period
distribution for the 43 binaries in our sample compared to the
corresponding distribution for the field, on a logarithmic scale. The
normalization was computed by requiring that the integral of the
smooth curve up to periods of 1~yr (our completeness level,
represented with a dashed line) is equal to the number of binaries
detected in our sample up to that period (40).  The hatched area of
the histogram represents the double-lined systems, and the open area
added above is for the SB1 binaries. 

\placefigure{figpdist}
	
The observed distribution of X-ray binaries is sharply peaked at short
periods (median $\log P = 0.67$, or a period of 4.7~days) and shows an
excess over the field distribution, as anticipated above. The
frequency of binaries in our sample (40 systems detected up to our
completeness limit of 1~yr, out of 121 objects studied) is
33\%~$\pm$~5\%. This is nearly double the fraction of binaries in the
field up to this period, which is 17\%~$\pm$~3\%. If we choose a more
conservative completeness limit for our sample of, say, 100~days, the
excess of binaries would be even greater since the number of systems
detected in our sample up to 100~days remains the same, yet the binary
frequency in the field decreases to 11\%~$\pm$~2\%. 

The increased fraction of binaries compared to the field is understood
as a selection effect, since binaries will tend to be brighter than
single stars in X-rays, not only because they are composed of two
objects, but also because of the increased activity if the period is
short. They will thus tend to be promoted into the (flux-limited)
sample more easily compared to single stars. A similar effect can
explain why SB2 systems dominate over SB1 systems. 

\subsection{Activity}
\label{secactivity}

Correlations between some of the physical properties of the stars and
a number of activity indicators might in principle be expected to
exist in a sample such as ours, composed largely of RS~CVn-type or
similarly active systems. For example, we searched for a dependence of
the X-ray hardness ratios HR1 and HR2 and also the X-ray flux of our
binaries (using data as reported by \markcite{m97}M97) with parameters
such as the orbital period, the effective temperature, the projected
rotational velocity, and the surface gravity. No significant
correlations were found, possibly because essentially all of our
objects are active by virtue of their selection. For the double-lined
systems where there is ambiguity as to which component's parameters to
use, we considered alternatively those of the coolest component, the
more massive component, and also the visually brightest component,
with null results in all cases. No differences were found between the
X-ray properties of the single-lined binaries and the double-lined
binaries, with the possible exception of a hint that the SB2 systems
may have systematically larger HR2 indices (by about 0.10-0.20),
indicative of slightly harder X-ray emission. 

\begin{figure}
 \plotfiddle{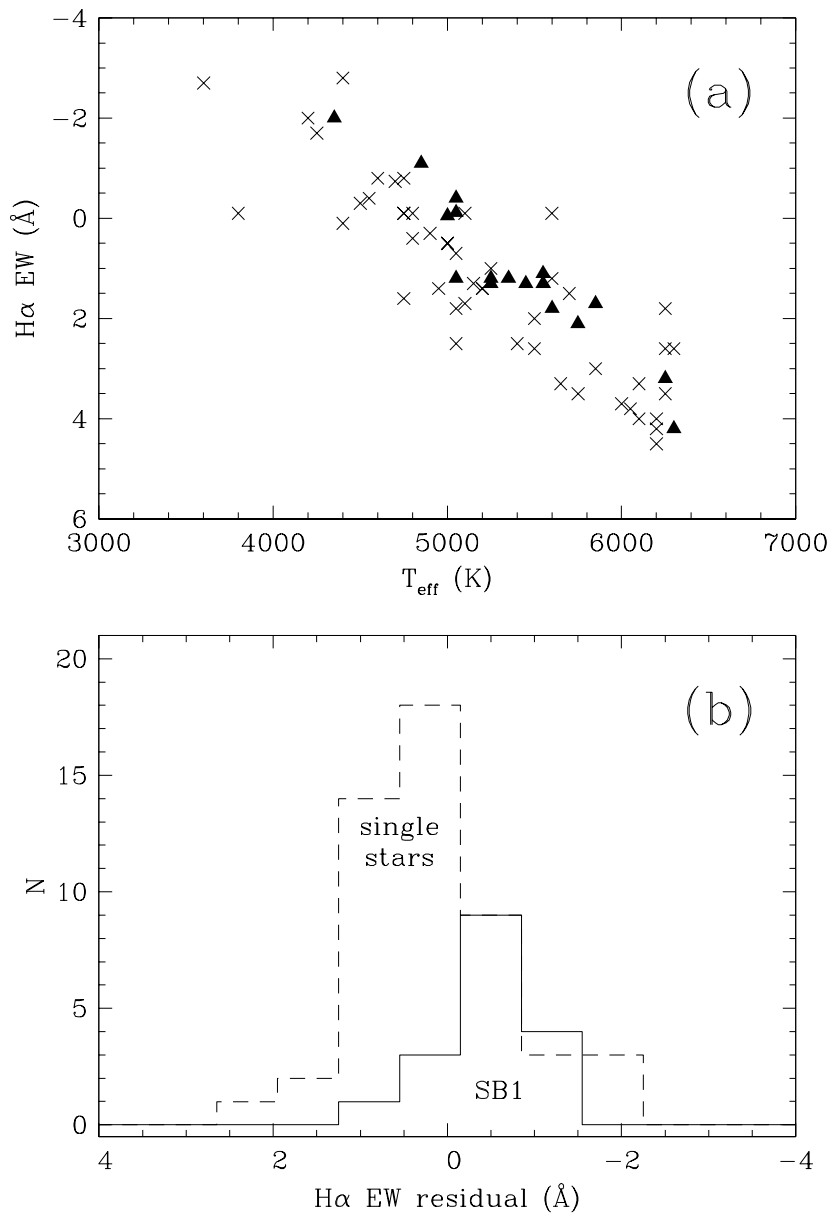}{6.5in}{0}{75}{75}{-240}{-50}
\figcaption[Torres.fig09.ps]{(a)~H$\alpha$ equivalent width (\AA) as a
function of the effective temperature, separately for the single-lined
binaries in the present sample (triangles) and the single stars in N97
(crosses). Negative values indicate emission, and positive values
represent absorption (partially filled-in or not). A similar
correlation is seen for both kinds of objects. (b)~Histogram of
residuals from a linear fit to all points in (a), to highlight the
fact that the binaries tend to have slightly stronger emission than
the single stars (see text).\label{fighalpha}}
 \end{figure}

Similarly, examination of correlations between the H$\alpha$ emission
as given by \markcite{m97}M97 (equivalent width) and the orbital or
stellar characteristics $\log P$, $v \sin i$, or $\log g$ indicated no
dependence. There is, however, a clear dependence with effective
temperature, in the sense that cooler systems tend to have stronger
H$\alpha$ emission. This is consistent with the fact that stars with
lower temperatures are typically more chromospherically active.  The
correlation is shown in Figure~\ref{fighalpha}a for the single-lined
systems in our sample (triangles).  Negative equivalent widths
indicate emission, while positive values represent absorption (which
may be partially filled-in in some cases).  Although there appears to
be a similar correlation for the double-lined systems (not shown), the
scatter is much larger because of the confusion as to which component
the emission measured by \markcite{m97}M97 corresponds to. 

\placefigure{fighalpha}

It is of interest to compare the trend for the single-lined binaries
with that for the single stars in the \markcite{n97}N97 sample, which
we have represented by crosses in Figure~\ref{fighalpha}a. The
temperatures for these single stars were derived using the same
technique we have applied in this paper.  The increase in H$\alpha$
equivalent width for the cooler objects is again obvious. The slope of
the correlation is essentially the same in both cases, but the
binaries appear to lie slightly higher on the diagram. To test the
statistical significance of this we carried out a linear fit to all
the data (SB1 $+$ singles) and examined the residuals from this fit
separately for the SB1 systems and the single stars. The slope of this
correlation is $d({\rm H}\alpha)/dT_{\rm eff} = +0.236 \pm 0.026$~\AA\
per 100~K.  Figure~\ref{fighalpha}b shows the distributions of
residuals. The SB1 systems have equivalent widths that are $0.56 \pm
0.18$~\AA\ more negative, on average, indicating stronger emission (a
3.1$\sigma$ effect). 

\begin{figure}
 \plotfiddle{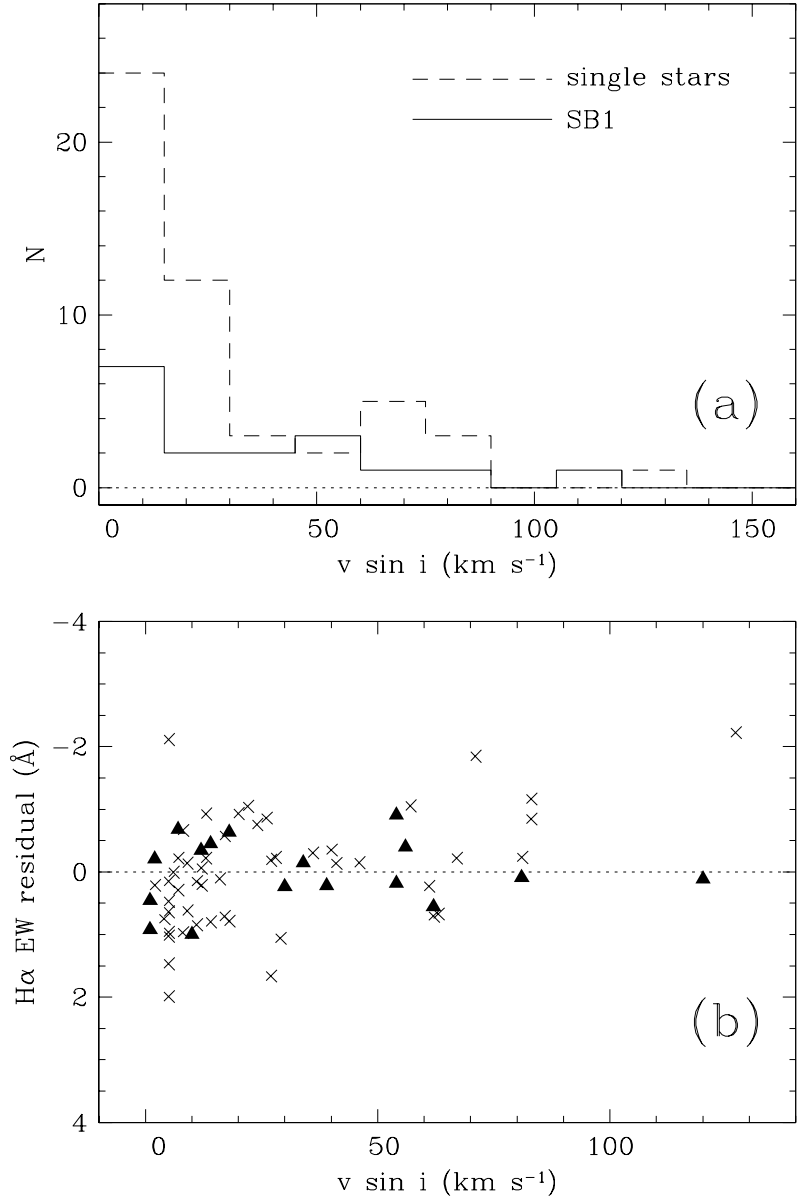}{6.5in}{0}{75}{75}{-240}{-50}
\figcaption[Torres.fig10.ps]{(a)~Distribution of the projected rotational
velocities for the single stars in N97 and for the single-lined
binaries in our sample. No significant differences are seen.
(b)~Residuals from a linear fit to Figure~\ref{fighalpha}a, corrected
for the systematic offset shown in Figure~\ref{fighalpha}b, separately
for the SB1s and the single stars. Symbols as in
Figure~\ref{fighalpha}a.\label{figvsini}}
 \end{figure}

To examine whether this excess emission compared to the single stars
might be due to a difference in rotational velocities, we show in
Figure~\ref{figvsini}a the $v \sin i$ distributions for the single
stars and for the SB1s. Despite expectations (tidal locking in close
binaries), the distributions are not statistically different, as
indicated by the Kolmogorov-Smirnov test. As it turns out, a
significant fraction of the binaries have long orbital periods, and
are therefore not yet affected by tidal effects. The reason for the
slightly stronger H$\alpha$ emission for the binaries in our sample is
thus not obvious. 

\placefigure{figvsini}

The vertical scatter in Figure~\ref{fighalpha}a, particularly for the
single stars, does appear to be related to the rotation of the stars.
In Figure~\ref{figvsini}b we show the residuals from the H$\alpha$
vs.\ $T_{\rm eff}$ correlation as a function of the rotational
velocity of the objects, separately for the single stars and the
binaries.  The residuals for both types of objects have been corrected
for the systematic difference found above. The trend for the single
stars goes in the direction expected, namely, the fast rotators show
extra emission and the slow rotators show less.  There appears to be
no such trend for the single-lined binaries, although the sample is
small. 
	
\section{Final remarks}

High-resolution spectroscopic monitoring of ROSAT X-ray sources south
of the Taurus-Auriga star-forming region has resulted in the discovery
of numerous binary systems in the \markcite{n97}N97 sample.  In this
paper we report new spectroscopic orbital solutions for 42 systems,
along with determinations of the physical characteristics (effective
temperature, projected rotational velocity, surface gravity, estimated
masses and radii) of all visible components.  Detailed descriptions of
each system have been provided to facilitate follow-up in particularly
interesting cases. 

The original selection of the \markcite{n97}N97 sample was designed to
increase the likelihood of including young stars associated with the
star-forming region, but at least in the case of the binaries that we
have turned up, it has generally yielded field objects of a rather
different nature than intended, ranging from contact systems of the
W~UMa type and mass-exchange Algols to active main sequence binaries.
The overall properties of these systems (strong excess of short-period
binaries, large fraction of double-lined systems) as well as the high
frequency of binaries compared to the number of original targets in
the sample can all be understood in terms of the selection criteria,
since these properties all tend to make the objects brighter in
X-rays. The finding that a significant fraction of the ROSAT X-ray
sources are active binaries is consistent with results in other
regions of the sky, including old open clusters such as M67 (see,
e.g., \markcite{b93}Belloni, Verbunt, \& Schmitt 1993;
\markcite{b98}Belloni, Verbunt, \& Mathieu 1998; \markcite{vdb99}van
den Berg, Verbunt, \& Mathieu 1999), even though the biases may be
somewhat different.  It seems likely, therefore, that surveys of ROSAT
X-ray sources in other areas of the sky using similar criteria to
promote the detection of PMS stars will also result in a possibly
substantial fraction of the objects being active binary systems
instead. However, that fraction could depend strongly on the region or
on the distance from the molecular gas. 

While most of our binary systems fall into the RS~CVn category, the
selection has at least been successful in two cases where other
evidence (\ion{Li}{1}~$\lambda$6708 absorption, kinematics, proximity
to other young objects, dynamical evidence based on our orbital
solutions) shows quite convincingly that the objects are truly young.
[41]~RXJ0528.9$+$1046 is a double-lined binary with a period of 7.7
days that is most likely associated with the $\lambda$~Ori region, and
may be as young as 2~Myr.  [42]~RXJ0529.3$+$1210 is a very eccentric
single-lined binary with an orbital period of 463 days, also a
probable member of the same region.  Both are classified as wTTS based
on their weak H$\alpha$ emission.  These systems add to the still
relatively small population of PMS binaries with known orbital
elements, the study of which may hold important clues to the process
of binary star formation. 

Among other interesting systems we have discovered is a potentially
important eclipsing binary, [6]~RXJ0239.1$-$1028, composed of a pair
of detached K stars. Such objects are relatively rare, and are
particularly valuable to test models of stellar evolution in the lower
main sequence. Three other systems in our sample are also very likely
to be eclipsing, and should be followed up: [3]~RXJ0212.3$-$1330,
[12]~RXJ0309.1$+$0324N, and [31]~RXJ0442.3$+$0118. The second of these
is a W~UMa-type binary. 

Finally, the short-period systems in our sample with circular orbits
have provided us with the opportunity to apply a simple but effective
tool for the study of double-lined binaries in general, and young
systems in particular, which is often overlooked: the use of the
stellar density as a sensitive measure of evolution. Under favorable
conditions the density can be derived directly from the spectroscopic
orbital elements and the measured projected rotational velocities, and
can serve as a substitute for the luminosity in the familiar H-R
diagram, with the advantage of being completely independent of
distance. The experience gained from the systems studied here shows
that it can be a useful diagnostic in some cases for distinguishing
between main sequence and PMS status, and for comparing with stellar
evolution models. 
	
\acknowledgments

Many of the spectroscopic observations for this project were obtained
at the CfA telescopes by P.\ Berlind, M.\ Calkins, J.\ Caruso, R.\ J.\
Davis, J.\ Degnan, D.\ W.\ Latham, A.\ A.\ E.\ Milone, J.\ Peters, R.\
P.\ Stefanik, and J.\ Zajac. We are grateful to R.\ J.\ Davis for also
maintaining the CfA database of radial velocities, and to Sabine
Frink, who provided proper motion information for some of our targets.
A number of helpful comments by an anonymous referee are also
acknowledged.  This research has made use of the SIMBAD database,
operated at CDS, Strasbourg, France, and of NASA's Astrophysics Data
System Bibliographic Services.

\clearpage

\clearpage

\clearpage

\clearpage

\clearpage

\begin{figure}
\plotfiddle{Torres.fig04b.ps}{7.34in}{0}{80}{80}{-260}{-40}
\end{figure}

\clearpage

\clearpage

\clearpage

\clearpage

\clearpage

\clearpage

\clearpage

\begin{figure}
 \plotfiddle{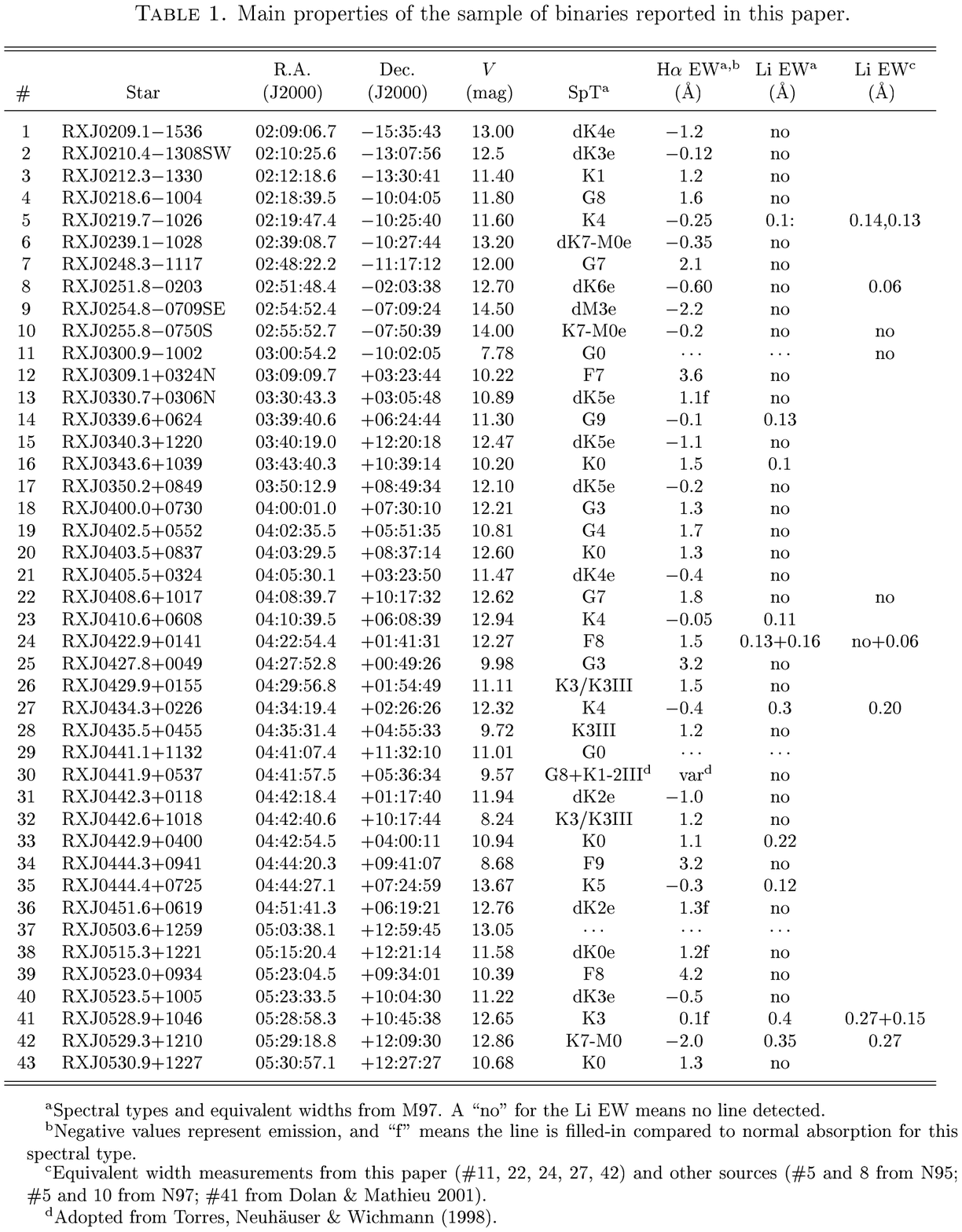}{6.5in}{0}{90}{90}{-270}{-90}
\end{figure}

\clearpage
	
\begin{deluxetable}{crrc}
\tablecolumns{4}
\tablewidth{23pc}
\tablenum{2}
\tablecaption{Measured (heliocentric) radial velocities for the
single-lined binaries in the sample.}
\tablehead{
\colhead{HJD} & \colhead{RV$_A$} & \colhead{(O$-$C)$_A$} & \colhead{Orbital} \\
\colhead{(2,400,000$+$)} & \colhead{($\kms$)} & \colhead{($\kms$)} & \colhead{phase}
}
\startdata
\multicolumn{4}{l}{[2] RXJ0210.4$-$1308SW   (02:10:25.6 $-$13:07:56)} \\
~~~49732.5746 &    53.93  &  $-$0.77  &   0.912 \\
~~~50033.7571 &    10.31  &  $-$2.10  &   0.271 \\
~~~50052.6145 &  $-$9.38  &  $-$3.36  &   0.655 \\
~~~50089.6196 & $-$15.73  &  $-$0.56  &   0.391 \\
~~~50093.6267 &    28.60  &  $-$0.13  &   0.210 \\
\enddata
\tablecomments{The complete version of this table is in the electronic
edition of the Journal.  The printed edition contains only a sample.}
\label{tabsb1rvs}
\end{deluxetable}

\begin{deluxetable}{crrrrc}
\tablecolumns{6}
\tablewidth{33pc}
\tablenum{3}
\tablecaption{Measured (heliocentric) radial velocities for the
double-lined binaries in the sample.}
\tablehead{
\colhead{HJD} & \colhead{RV$_A$} & \colhead{RV$_B$} & \colhead{(O$-$C)$_A$} & 
\colhead{(O$-$C)$_B$} & \colhead{Orbital} \\
\colhead{(2,400,000$+$)} & \colhead{($\kms$)} & \colhead{($\kms$)} & 
\colhead{($\kms$)} & \colhead{($\kms$)} & \colhead{phase}
}
\startdata
\multicolumn{6}{l}{[1] RXJ0209.1$-$1536     (02:09:06.7 $-$15:35:43)} \\
~~~49732.5908  &   16.13  &    27.62  & $-$2.87  & $+$13.52  &   0.590 \\
~~~49964.8905  &   87.30  & $-$71.50  & $-$2.80  &  $-$2.21  &   0.850 \\
~~~50002.8802  & $-$8.99  &    56.00  & $+$2.49  &  $+$6.15  &   0.519 \\
~~~50004.8241  &   90.37  & $-$82.45  & $+$0.54  & $-$13.48  &   0.856 \\
~~~50033.7465  &   36.56  &  $-$4.19  & $+$2.02  &  $-$0.07  &   0.625 \\
\enddata
\tablecomments{The complete version of this table is in the electronic
edition of the Journal.  The printed edition contains only a sample.}
\label{tabsb2rvs}
\end{deluxetable}

\clearpage

\begin{figure}
 \plotfiddle{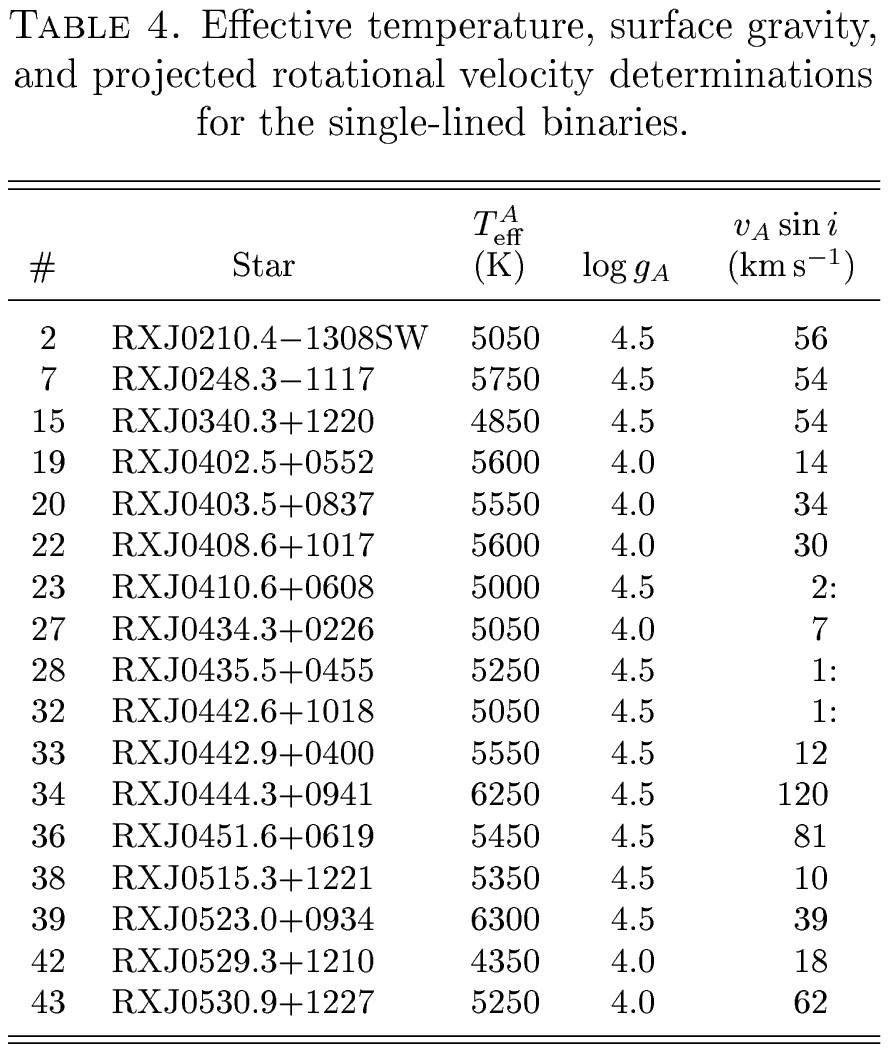}{6.5in}{0}{100}{100}{-310}{-110}
\end{figure}

\clearpage

\begin{figure}
 \plotfiddle{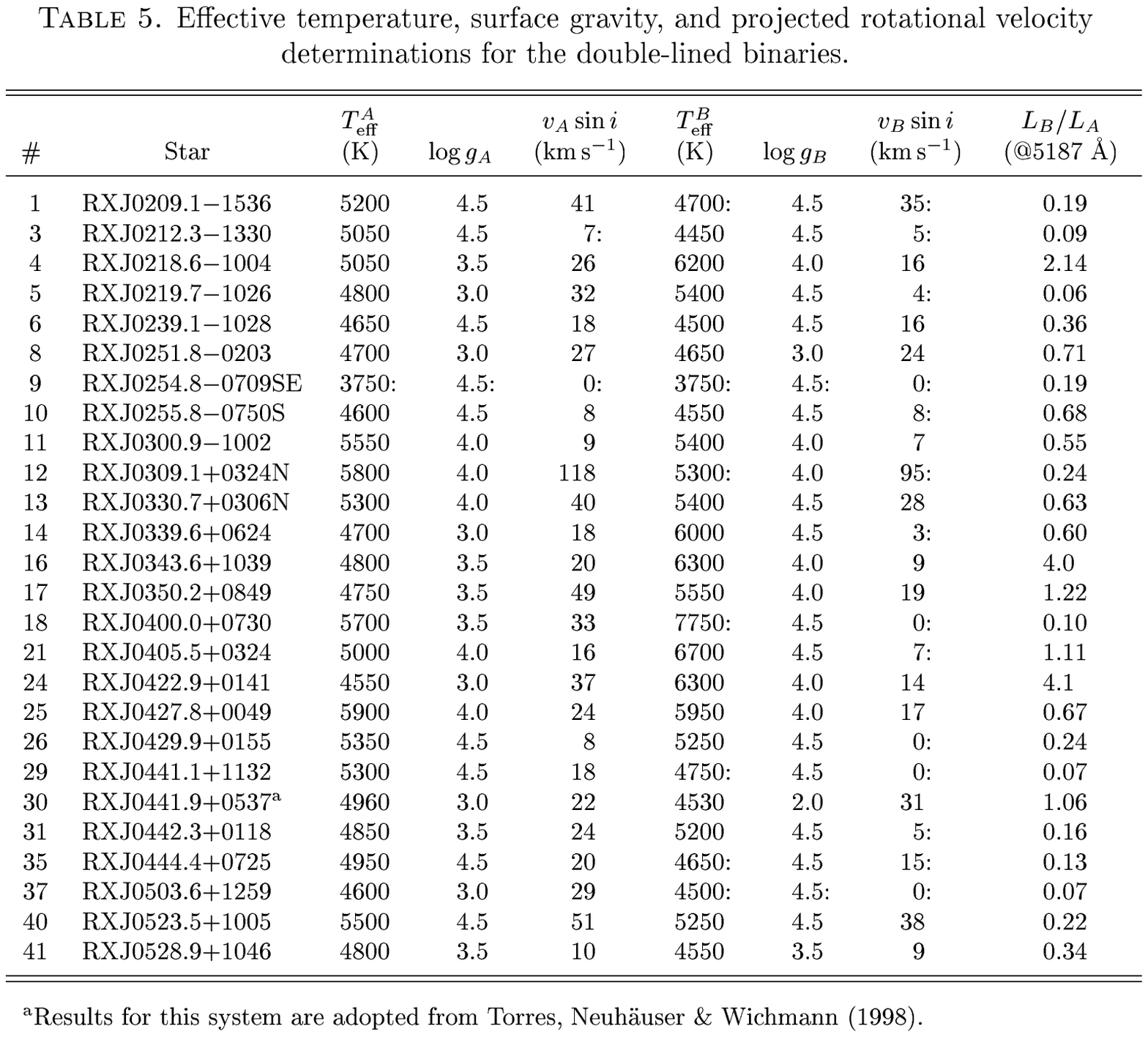}{6.5in}{0}{100}{100}{-310}{-130}
\end{figure}

\clearpage

\begin{figure}
 \plotfiddle{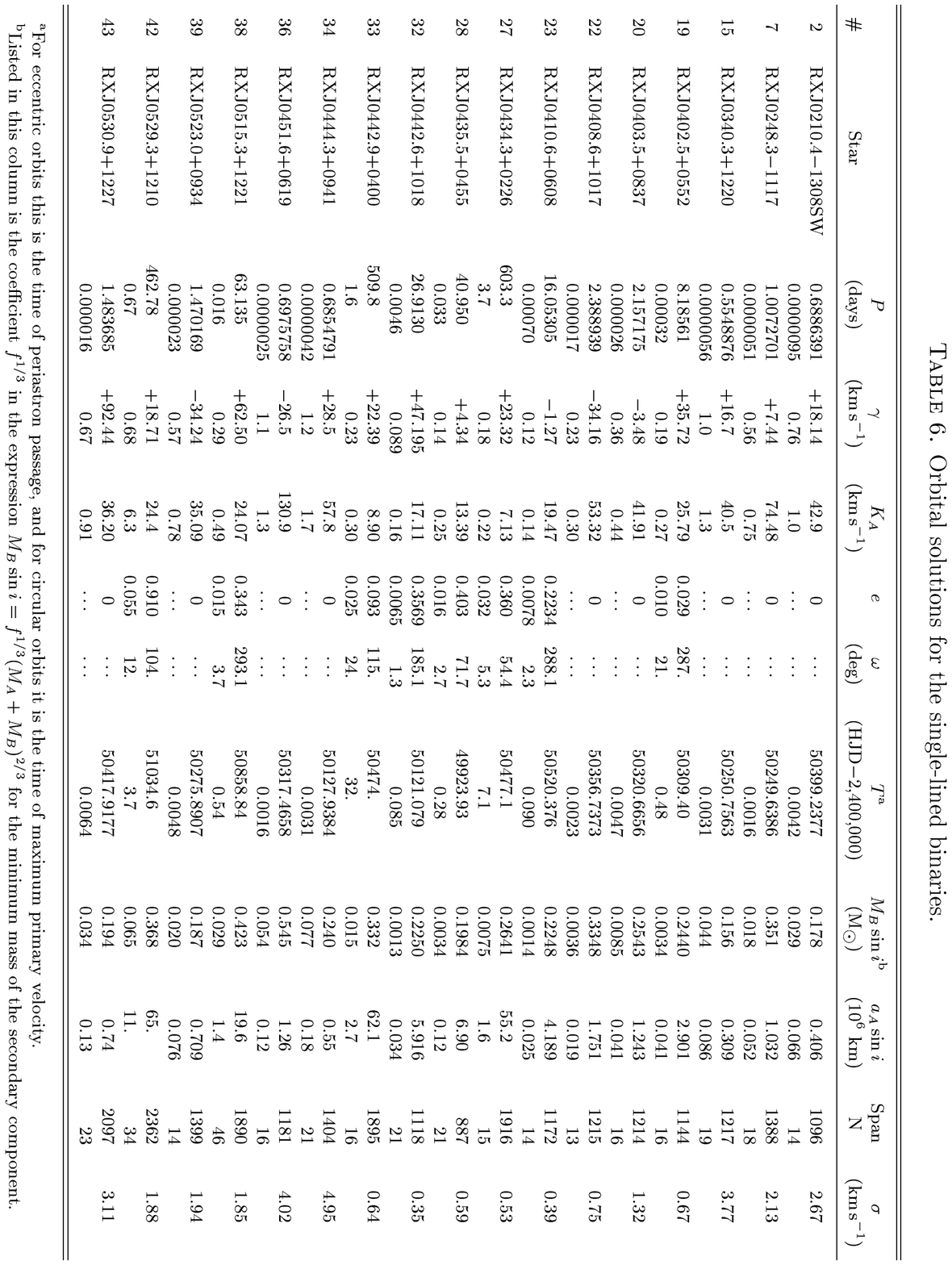}{6.5in}{180}{95}{95}{290}{630}
\end{figure}

\clearpage

\begin{figure}
 \plotfiddle{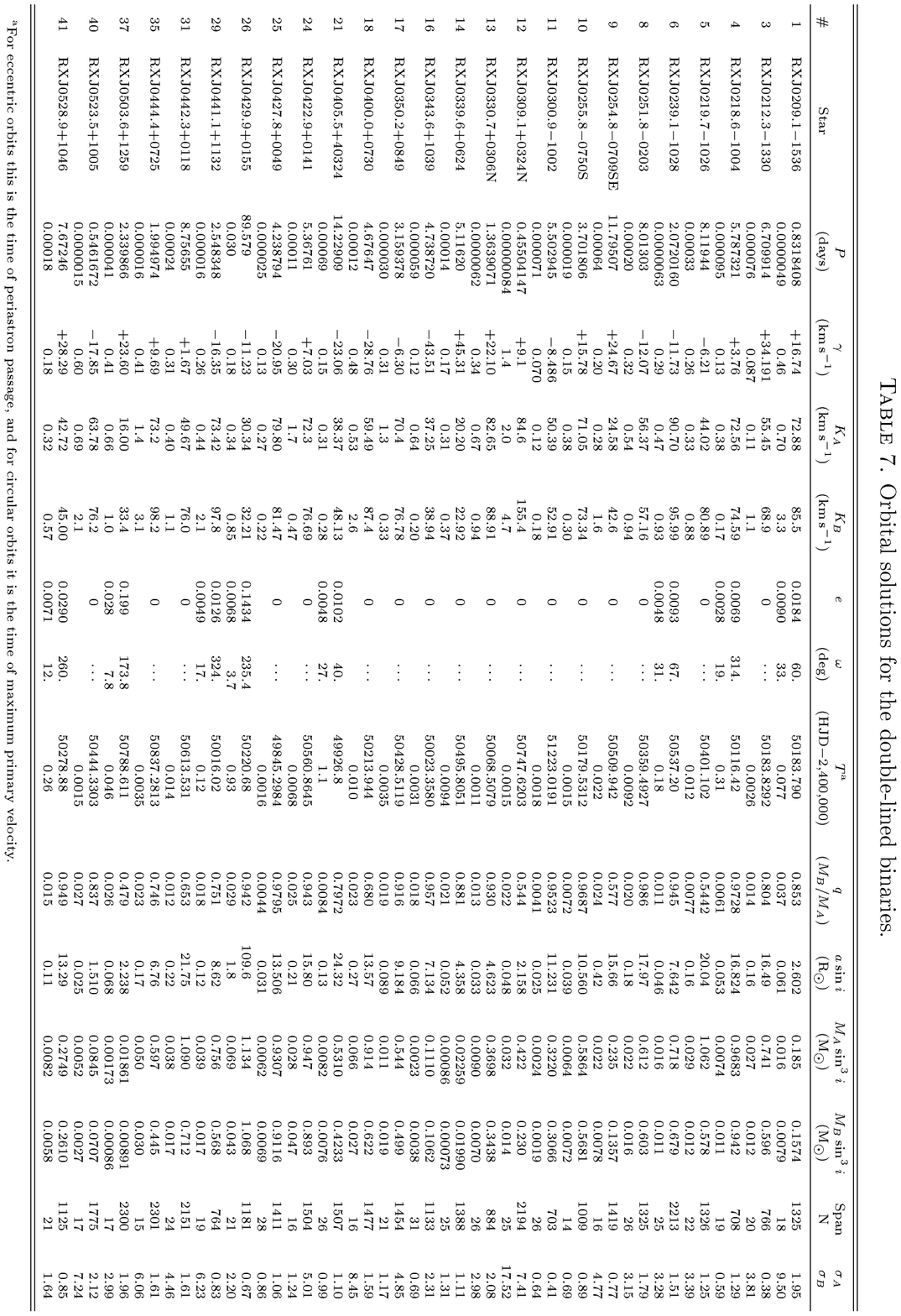}{6.5in}{180}{90}{90}{280}{590}
\end{figure}

\end{document}